\documentclass[camera]{jpaper}
\usepackage[sort]{cite}

\usepackage{soul}

\usepackage{amsmath}
\usepackage{graphicx}
\usepackage[linesnumbered,ruled]{algorithm2e}

\usepackage{algpseudocode}
\usepackage{titlesec}
\algrenewcommand\algorithmiccomment[2][\normalsize]{{#1\hfill\(\triangleright\) #2}}
\usepackage{geometry} \usepackage{array}
\makeatletter
\let\MYcaption\@makecaption
\makeatother

\usepackage[font=footnotesize]{subcaption}

\makeatletter
\let\@makecaption\MYcaption
\makeatother

\usepackage{fixltx2e}

\usepackage{dblfloatfix}
\usepackage[nolessnomore, italic]{mathastext}
\usepackage[T1]{fontenc}
\usepackage[usenames,dvipsnames,svgnames,table]{xcolor}
\usepackage{multirow}
\usepackage{hhline}
\usepackage[normalem]{ulem}
\usepackage{setspace}
\usepackage{indentfirst}
\usepackage[hang,flushmargin]{footmisc}
\usepackage{graphicx}

\usepackage{comment}
\usepackage{amssymb} 
\usepackage{xcolor}
\usepackage{textcomp}
\usepackage{fancyhdr}
\usepackage{microtype}
\usepackage{url}
\usepackage{flushend}
\usepackage{ulem}
\usepackage[bookmarks=true,breaklinks=true,letterpaper=true,colorlinks,linkcolor=black,citecolor=blue,urlcolor=black]{hyperref}
\usepackage{setspace}
\usepackage{mdframed} 
\usepackage{mathptmx} 
\usepackage{enumitem}
\usepackage{tikz}
\usepackage{subcaption} 
\usepackage{booktabs} 
\usepackage{caption}
\usepackage{dblfloatfix}
\usepackage{multirow} 
\usepackage{float}
\usepackage{color}

\usepackage{xcolor}
\definecolor{darkgreen}{RGB}{0,50,0}
\definecolor{olivegreen}{RGB}{68,85,37}
\definecolor{darkred}{RGB}{190,0,0}
\definecolor{darkblue}{RGB}{0,0,190}
\definecolor{mygreen}{RGB}{2,108,69}

\usepackage{listings}

\usepackage{color}
\definecolor{codegray}{rgb}{0.5,0.5,0.5}
\lstdefinestyle{mystyle}{
    keywordstyle=\color{blue}\textbf,
    numberstyle=\footnotesize\color{codegray},
    basicstyle=\footnotesize\fontfamily{txtt}\selectfont,
    breakatwhitespace=false,
    breaklines=true,
    captionpos=b,
    keepspaces=true,
    numbers=left,
    numbersep=5pt,
    showspaces=false,
    showstringspaces=false,
    showtabs=false,
    tabsize=2,
    firstnumber=last
}
\usepackage{paralist}

\definecolor{blue(pigment)}{rgb}{0.2, 0.2, 0.7}
\definecolor{dgreen}{rgb}{0.0, 0.8, 0.0}
\usepackage{xcolor}
\definecolor{ocre}{RGB}{243,102,25}
\newmdenv[%
linecolor=blue(pigment),
backgroundcolor=blue(pigment)!10,
linewidth=0pt]{mytablebox}

\definecolor{amber}{rgb}{1.0, 0.49, 0.0}
\definecolor{darkbyzantium}{rgb}{0.36, 0.22, 0.33}
\definecolor{darkseagreen}{rgb}{0.56, 0.74, 0.56}
\definecolor{darkspringgreen}{rgb}{0.09, 0.45, 0.27}
\definecolor{dollarbill}{rgb}{0.52, 0.73, 0.4}
\definecolor{MidnightBlue}{rgb}{0.1, 0.1, 0.44}

\definecolor{darkgreen}{RGB}{0,50,0}
\definecolor{olivegreen}{RGB}{68,85,37}
\definecolor{darkred}{RGB}{190,0,0}
\definecolor{darkblue}{RGB}{0,0,190}

\newcommand\arxiv[1]{\noindent{\color{black}{#1}}} 
\newcommand\tofix[1]{\noindent{\color{amber}{#1}}} 

\newcommand{\smartpq}{\emph{SmartPQ}}
\newcommand{\numa}{\emph{NUMA-aware}}

\newcommand{\notnuma}{\emph{NUMA-oblivious}}
\newcommand{\insrt}{\emph{insert}} 
\newcommand{\delete}{\emph{deleteMin}}
\newcommand{\nuddle}{\emph{Nuddle}}
\newcommand{\algomode}{algorithmic}
\newcommand{\ffwd}{\emph{ffwd}}

\newcommand\blfootnote[1]{%
  \begingroup
  \renewcommand\thefootnote{}\footnote{#1}%
  \addtocounter{footnote}{-1}%
  \endgroup
}

\newcommand{\affilNTUA}[0]{\textsuperscript{$\dagger$}}

\newcommand{\affilETH}[0]{\textsuperscript{$\ddagger$}}

\usepackage{pifont}

\begin{document}
\bstctlcite{IEEEexample:BSTcontrol} 

\title{\textbf{\smartpq}: An Adaptive Concurrent Priority Queue \\
for NUMA Architectures\vspace{-6pt}}

%


\author{
\hspace{-12pt}
\fontsize{11.4}{8}\selectfont
\parbox[t]{1.02\textwidth}{
{\text{*}Christina Giannoula\affilNTUA}\hspace{12pt}
{\text{*}Foteini Strati\affilETH\affilNTUA}\hspace{12pt}
{Dimitrios Siakavaras\affilNTUA}\hspace{12pt}
{Georgios Goumas\affilNTUA}\hspace{12pt}
{Nectarios Koziris\affilNTUA}
{\vspace{4pt}\textcolor{white}{}}
\\
\centering
\emph{{\affilNTUA National Technical University of Athens\qquad \affilETH ETH Z{\"u}rich 
}}%
}
\vspace{-4pt}
}

\maketitle

\thispagestyle{plain} 
\pagestyle{plain}




\fancyhead{}

\setstretch{0.944}




\begin{abstract}

Concurrent \arxiv{priority queues are widely used in important workloads, such as graph applications and discrete event simulations. However, } designing scalable concurrent priority queues for NUMA \arxiv{architectures} is challenging. \arxiv{Even though several} \notnuma{} implementations can scale up to a high number of threads\arxiv{,} exploiting the potential parallelism of \insrt{} operation\arxiv{,} \arxiv{\notnuma{} implementations scale poorly in \delete-dominated workloads. This is because all} threads compete for accessing the \emph{same} memory locations, i.e.\arxiv{,} the highest-priority \arxiv{element of} the queue\arxiv{, thus \arxiv{incurring} excessive cache coherence traffic and non-uniform memory accesses \emph{between NUMA} nodes.} In such \arxiv{scenarios}, \numa{} implementations are typically used \arxiv{to improve system performance on} a NUMA system.\blfootnote{\textbf{* Christina Giannoula and Foteini Strati are co-primary authors.}}

In this work, we propose an adaptive priority queue, called \smartpq. \smartpq{} tunes itself by switching between a \notnuma{} and a \numa{} \algomode{} mode to \arxiv{achieve high performance under} \arxiv{all various contention scenarios}. \arxiv{\smartpq{} has two key components. First,} it is built on top of NUMA Node Delegation (\nuddle), a generic low-overhead technique to construct \arxiv{efficient} \numa{} data structures using any arbitrary \arxiv{concurrent} \notnuma{} implementation as its backbone. \arxiv{Second}, \smartpq{} \arxiv{integrates a lightweight decision making mechanism} to decide when to switch between \arxiv{\notnuma{} and \numa{}} \algomode{} modes. Our evaluation \arxiv{shows that, in NUMA systems, \smartpq{}} \arxiv{performs best in all various contention scenarios} with 87.9\% success rate, and dynamically adapts between \numa{} and \notnuma{} \algomode{} mode, \arxiv{with negligible performance} overheads. \arxiv{\smartpq{} improves performance by} \arxiv{1.87$\times$ on average over} SprayList, the state-of-the-art \notnuma{} priority queue.

\end{abstract}

\section{Introduction}
\label{Introductionbl}

Concurrent \arxiv{data structures are widely used in the software stack, i.e., kernel, libraries and applications. Prior works~\cite{ffwd,blackbox,numask,adaptivepq} discuss the need for efficient and scalable concurrent data structures for commodity Non-Uniform Memory Access (NUMA) architectures.} Pointer chasing data structures such as linked lists, skip lists and search trees have inherently \arxiv{low contention}, \arxiv{since concurrent threads search for different elements during their operations}. Recent works~\cite{blackbox,ascylib,Siakavaras2021RCUHTM} have shown that lock-free algorithms ~\cite{Harris-ll,Michael-ll, fraser, Ellen-bst, Howley-bst, Natarajan-bst} of such data structures can scale to hundreds of cores. On the other hand, data structures such as queues and stacks \arxiv{typically incur high contention,} when accessed by many threads. In these data structures, concurrent threads compete for the \emph{same} memory elements (locations), \arxiv{incurring} excessive traffic \arxiv{and non-uniform memory accesses} between nodes of a NUMA system.

In this work, we focus on priority queues, which are widely used in a \arxiv{variety} of applications, including task scheduling in real-time and computing systems~\cite{Xu}, discrete event simulations~\cite{Tang,Marotta} and graph \arxiv{applications}~\cite{Kolmogorov,Lasalle,Thorup}, \arxiv{e.g.,} Single Source Shortest Path~\cite{CLRS} and Minimum Spanning Tree~\cite{Prim}. \arxiv{Similarly to skip-lists and search trees, priority queues have two main operations: \insrt{} and \delete. The \insrt{} operation, concurrent priority queues typically exhibits high levels of parallelism and low-contention, since threads may work on different parts of the data structure. Therefore, concurrent \notnuma{} implementations~\cite{lotan_shavit, linden_jonsson, sagonas, sundell, Wimmer_Martin, rihani, Brodal, Zhang} can scale up to a high number of threads. In contrast,} in \delete{} operation, \emph{all} threads compete for \arxiv{deleting} the \arxiv{highest-priority} element of the queue, \arxiv{thus competing for the \emph{same} memory locations (similarly to queues and stacks), and} creating a contention spot. \arxiv{In \delete{}-dominated workloads, concurrent priority queues typically incur high-contention and low parallelism. To \arxiv{achieve higher} parallelism, \emph{relaxed} priority queues have been proposed in the literature~\cite{spraylist,Heidarshenas2020Snug}, in which \delete{} operation returns an element among the \emph{first few} (high-priority) elements of the priority queue. However, such \notnuma{} implementations are still inefficient in NUMA architectures, as we demonstrate in Section~\ref{sec:experimental}.} \arxiv{Therefore, to improve performance in NUMA systems}, \numa{} implementations have been proposed~\cite{blackbox,ffwd}.

\begin{figure}[t]
    \centering
    \includegraphics[scale=0.16]{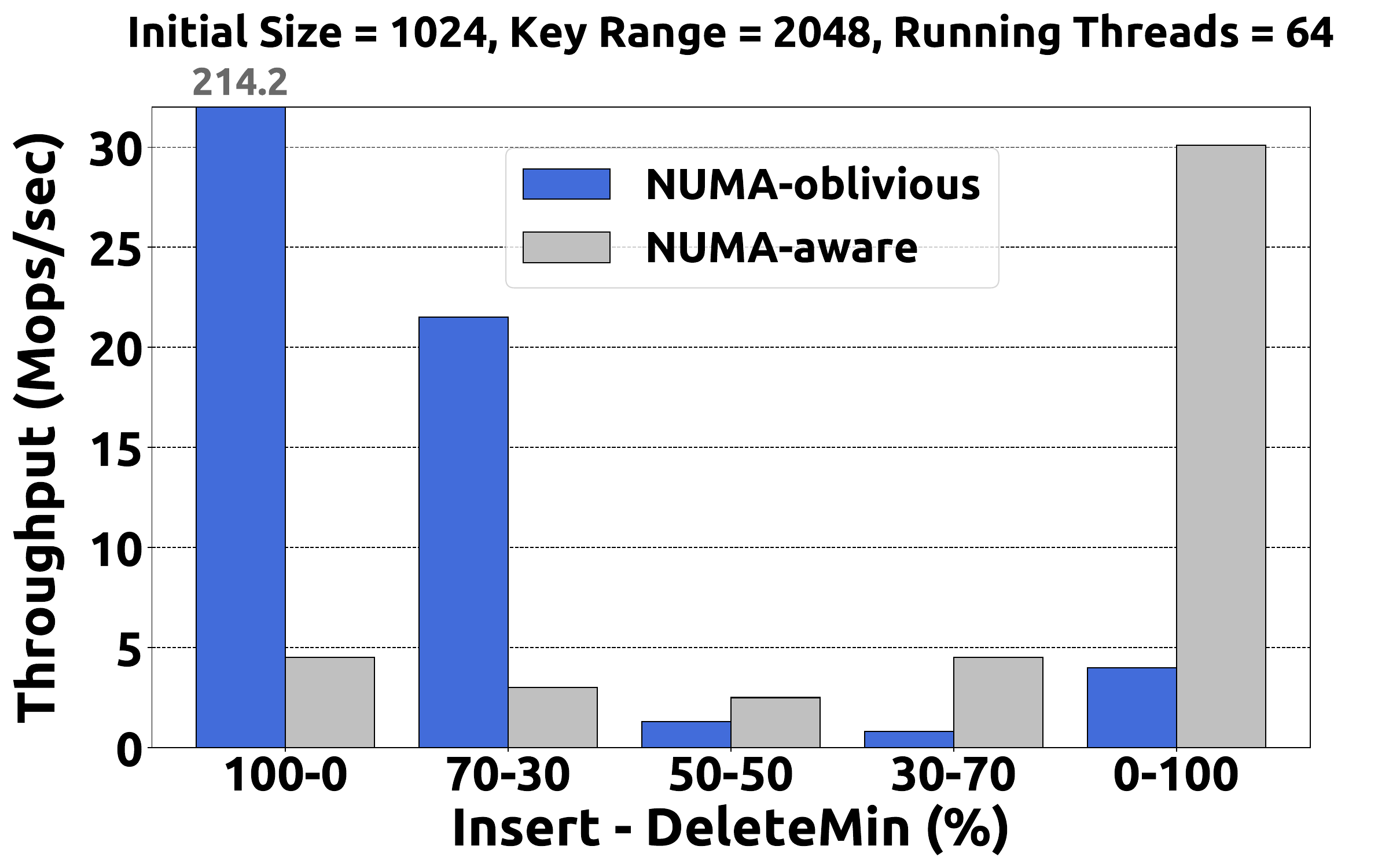}
    \caption{Throughput \arxiv{achieved} by a \notnuma{}~\cite{spraylist,herlihy} and a \numa{}~\cite{ffwd} priority queue, both initialized with 1024 keys. \arxiv{We use 64 threads that} perform a mix of \insrt{} and \delete{} operations in parallel, \arxiv{and the key range is set to 2048 keys.} We \arxiv{use \emph{all} NUMA nodes of} a 4-node NUMA system, the characteristics of which are presented in \arxiv{S}ection~\ref{sec:experimental}. }
    \label{fig:motivation}
\end{figure}

We examine \arxiv{\numa{} and \notnuma{} concurrent priority queues with a wide variety of contention scenarios in NUMA architectures, and find that the performance of a priority queue implementation is becoming increasingly dependent on both the contention levels of the workload and the underlying computing platform.} This is illustrated in \tofix{Figure~\ref{fig:motivation}}, which shows the throughput \arxiv{achieved by a} \notnuma{} and \arxiv{a} \numa{} priority queue \arxiv{using} a 4-node NUMA system. \arxiv{Even though in a \insrt-dominated scenario,} e.g., \arxiv{when having 100\% \insrt{} operations}, the \notnuma{} implementation \arxiv{achieves significant} performance \arxiv{gains over the \numa{} one}, when contention increases\arxiv{, i.e., the percentage of \delete{} operations increases, the \notnuma{} implementation incurs non-negligible performance slowdowns over the \numa{} priority queue. We conclude that none of the priority queues performs best across \emph{all} contention workloads.}


Our \textbf{goal} in this work is to design a concurrent priority queue that \arxiv{(i)} achieves \emph{the highest performance \arxiv{under all various contention scenarios}, and (ii) performs best} even when the \arxiv{contention of the workload \emph{varies} over time}.

To this end, our contribution is twofold. First, we introduce \textit{NUMA Node Delegation} (\nuddle{}), a generic technique \arxiv{to obtain \numa{} data structures, by effectively transforming \emph{any} concurrent \notnuma{} data structure into the corresponding \numa{} implementation. In other words, \nuddle{} is a framework to wrap any \emph{concurrent} \notnuma{} data structure} and transform it into an efficient \numa{} one. \nuddle{} extends \ffwd~\cite{ffwd} by \arxiv{enabling} multiple server threads, instead of only one, to execute operations \arxiv{in parallel} on behalf of client threads. In contrast to \ffwd, \arxiv{which aims to provide single threaded data structure performance,} \nuddle{} targets data structures \arxiv{which} are able to scale up to a number of threads \arxiv{such as priority queues.}



Second, we propose \smartpq, an adaptive \arxiv{concurrent} priority queue \arxiv{that achieves the \emph{highest} performance under \emph{all} contention workloads and \emph{dynamically} adapts itself over time between a \notnuma{} and a \numa{} \algomode{} mode. \smartpq{} integrates (i) \nuddle{} to \emph{efficiently} switch between the two \algomode{} modes with \emph{very low} overhead, and (ii) a simple decision tree \emph{classifier}, which predicts the best-performing \algomode{} mode} given the \arxiv{expected contention levels of a workload.}

\begin{figure}[t]
    \centering
    \includegraphics[scale=0.42]{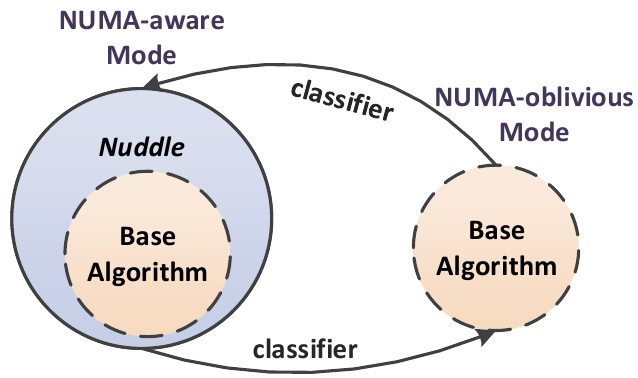}
    \caption{High-level \arxiv{overview} of \smartpq. \smartpq{} dynamically adapts its algorithm to the \arxiv{contention levels} of the workload \arxiv{based on} the prediction of \arxiv{a simple} classifier.}
\label{fig:smartpq}
\end{figure}

Figure~\ref{fig:smartpq} presents \arxiv{an overview of \smartpq}, where we use the term \textit{base algorithm} to denote \emph{any} arbitrary \arxiv{concurrent} \notnuma{} data structure. \arxiv{\smartpq{} relies on three key ideas. First,} client threads can execute operations using either \nuddle{} (\numa{} mode) or its underlying \notnuma{} \arxiv{base algorithm} (\notnuma{} mode). \arxiv{Second,} \smartpq{} incorporates a decision making mechanism to decide upon transitions between the two modes. \arxiv{Third, \smartpq{} exploits the fact that the actual underlying implementation of \nuddle{} is a \emph{concurrent} \notnuma{} data structure. Client threads in} both \algomode{} modes access the data structure \arxiv{with the same concurrency strategy}, \arxiv{i.e.,  with no actual change in the way data \arxiv{is} accessed, and synchronization is implemented. Therefore,} \smartpq{} switches from one mode to another with \emph{no} synchronization points between transitions.

We evaluate \arxiv{a wide range of contention scenarios and compare \nuddle{} and \smartpq{} with state-of-the-art \notnuma{}~\cite{spraylist,lotan_shavit} and \numa{}~\cite{ffwd} concurrent priority queues. We also evaluate \smartpq{} using synthetic benchmarks that \emph{dynamically} vary their contention workload over time.} Our evaluation \arxiv{shows that} \smartpq{} adapts between its two \algomode{} modes with \arxiv{negligible performance} overheads, and \arxiv{achieves} the highest performance \arxiv{in \emph{all} contention workloads
} with 87.9\% success rate.

This \arxiv{paper makes the following contributions:
\begin{compactitem}[$\textbf{--}$]
    \item We propose \nuddle{}, a generic technique to obtain \numa{} concurrent data structures.
    \item We design a simple classifier to predict the best-performing implementation among \notnuma{} and \numa{} priority queues given the contention levels of a workload.
    \item We propose \smartpq, an adaptive concurrent priority queue that achieves the highest performance, even when contention varies over time.
    \item We evaluate \nuddle{} and \smartpq{} with a wide variety of contention scenarios, and demonstrate that \smartpq{} performs best over prior state-of-the-art concurrent priority queues.
\end{compactitem}
} 

\section{NUMA Node Delegation (\nuddle)}
\label{sec:nuddle}

\subsection{Overview}
NUMA Node Delegation (\nuddle) is a generic technique \arxiv{to obtain \numa{} data structures by automatically transforming \emph{any} concurrent} \notnuma{} data structure into an efficient \numa{} implementation. \arxiv{\nuddle{}} extends \ffwd~\cite{ffwd}, a client-server software mechanism \arxiv{which is based on the delegation technique~\cite{Calciu2013Message,Klaftenegger2014Delegation,Lozi2012Remote,Petrovic2015Performance,Suleman2009Accelerating}.}

\begin{figure}[t]
    \centering
    \hspace{-8pt}\includegraphics[scale=0.38]{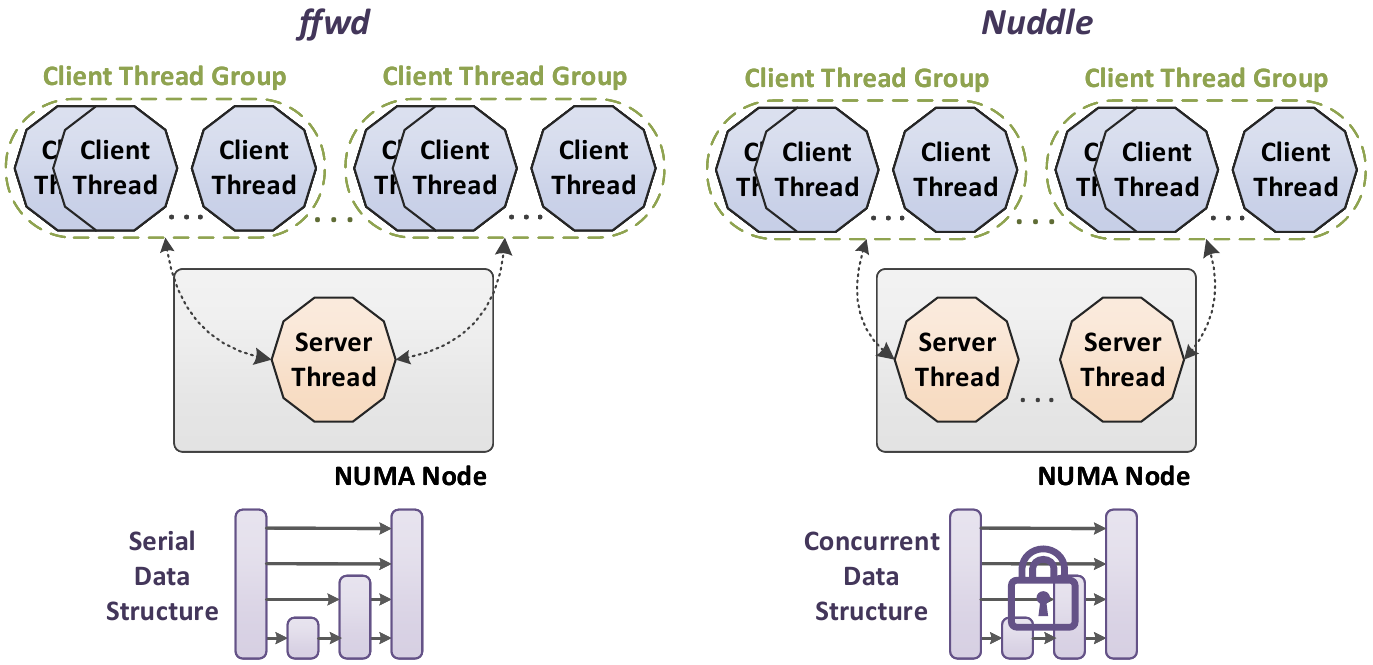}
    \vspace{1pt}
    \caption{High-level design of \ffwd~\cite{ffwd} and \nuddle{}. \nuddle{} \arxiv{locates \emph{all} server threads} at the \emph{same} NUMA node to \arxiv{design} a \numa{} \arxiv{scheme}, \arxiv{and associates} each of \arxiv{them to} multiple client \arxiv{thread} groups. \nuddle{} uses the communication protocol \arxiv{proposed in} \ffwd~\cite{ffwd}.}
    \label{fig:nuddle}
\end{figure}

\arxiv{Figure \ref{fig:nuddle} left shows the high-level overview of \ffwd, which has three key design characteristics. First,} \emph{all} operations performed by \arxiv{multiple} client threads are \emph{delegated} to \arxiv{one} \emph{single} dedicated thread, called \textit{server} \arxiv{thread}. \arxiv{Server thread performs operations in the data structure on behalf of its client threads. This way, the data structure remains in the memory hierarchy of a \emph{single} NUMA node, avoiding non-uniform memory accesses to remote data. Second,} \ffwd{} eliminates the need for synchronization, since the shared data structure is no longer accessed by multiple threads\arxiv{: only} a single server thread directly \arxiv{modifies} the data structure\arxiv{, and therefore,} \ffwd{} \arxiv{uses} a \emph{serial asynchronized} implementation of the underlying data structure. \arxiv{Third,  \ffwd{} provides an efficient communication protocol between the server thread and client threads that minimizes cache coherence overheads. Specifically, \ffwd{} reserves dedicated cache lines to exchange request and response messages between the client threads and sever thread.} Multiple client \arxiv{threads} are grouped together to minimize the response messages from the server thread\arxiv{: one response cache line is shared among multiple client threads belonging at the same client thread group.} For more details, \arxiv{we refer} the reader to \arxiv{the original paper}~\cite{ffwd}.

Figure \ref{fig:nuddle} right \arxiv{presents the high-level overview of \nuddle, \arxiv{which is based on three key ideas.} First, \nuddle{}} deploys \emph{multiple} servers \arxiv{to perform operations on behalf of multiple client threads. Specifically, client threads are grouped in client thread groups, and} each \arxiv{sever thread} serves multiple client \arxiv{thread} groups. \arxiv{This way, multiple server threads \emph{concurrently} perform operations on the data structure, achieving high levels of parallelism up to a number of server threads. Second, \nuddle{} locates all} server \arxiv{threads} to the \emph{same} NUMA node to keep the data structure \arxiv{in the memory hierarchy of one \emph{single} NUMA node, and propose a \numa{} approach. Client threads} can be located at any \arxiv{NUMA} node. \arxiv{Third, since} multiple servers \arxiv{can} \emph{concurrently} \arxiv{update} the \arxiv{\emph{shared}} data structure, \arxiv{\nuddle{} uses} the \emph{concurrent} \notnuma{} \arxiv{implementation (i.e., which includes synchronization primitives when accessing the shared data) of the underlying data structure} to ensure correctness. \arxiv{Third, \nuddle{} employs} the same client-server communication protocol with \ffwd{} \arxiv{to} carefully manage memory accesses \arxiv{and} minimize cache coherence traffic and latency.

\ffwd{} \arxiv{targets} inherently serial data structures, whose concurrent performance cannot be better than that of \emph{single} threaded \arxiv{performance. In contrast}, \nuddle{} targets data structures that can scale up to a number of concurrent threads. Priority \arxiv{queue is} a typical example of such a data structure. \arxiv{In \insrt{} operation, priority queue can scale up to multiple threads which can \emph{concurrently} update the shared data. In contrast, \delete{} operation is inherently serial: at each time only \emph{one} thread can update the shared data,} since \emph{all} threads \arxiv{compete for} 
the \arxiv{highest-priority} element \arxiv{of the queue}. \arxiv{However, as we mentioned, i}n relaxed priority queues (e.g., SprayList~\cite{spraylist}), even \delete{} \arxiv{operation} can be parallelized to some extent.

\subsection{Implementation Details}

Figures~\ref{alg:structs},~\ref{alg:init} and~\ref{alg:functions} present the code of a priority queue \arxiv{implementation using} \nuddle{}. We denote with red color the \arxiv{core} operations of the \arxiv{\emph{base algorithm}, which is used as the} underlying \arxiv{concurrent \notnuma{} implementation of \nuddle. Note that even} though in this work we focus on priority queues, \nuddle{} is a \arxiv{\emph{generic}} framework \arxiv{for any type of concurrent data structure.} 

\textbf{\arxiv{Helper S}tructures.} \arxiv{\nuddle{} includes three \emph{helper} structures (Figure~\ref{alg:structs}), which are needed for client-server communication. First,} the \arxiv{main} structure of \nuddle, \arxiv{called}  {\fontfamily{lmtt}\selectfont \textit{struct nuddle\_pq}}, wraps the \emph{base algorithm} ({\fontfamily{lmtt}\selectfont \textit{nm\_oblv\_set}}), and includes \arxiv{a few} additional fields\arxiv{, which are used} to associate client \arxiv{thread groups to} server \arxiv{threads} in the initialization step. \arxiv{Second, each} client \arxiv{thread} has its own {\fontfamily{lmtt}\selectfont \textit{struct client}} structure with a dedicated request and a dedicated response cache line. The request cache line is exclusively written by the client \arxiv{thread} and read by the \arxiv{associated} server \arxiv{thread}, while the response cache line is exclusively written by the server \arxiv{thread} and read by all client \arxiv{threads} that belong in the same client \arxiv{thread} group. \arxiv{Third}, each server \arxiv{thread} has its own {\fontfamily{lmtt}\selectfont \textit{struct server}} structure that includes an array of requests ({\fontfamily{lmtt}\selectfont \textit{my\_clients}}), each of them \arxiv{is} shared with a client \arxiv{thread}, and an array of responses ({\fontfamily{lmtt}\selectfont \textit{my\_responses}}), each of them \arxiv{is} shared with \arxiv{all client threads of the \emph{same}} client \arxiv{thread} group.

\begin{figure}[H]
\begin{lstlisting}[language=C]
#define cache_line_size 128
typedef char cache_line[cache_line_size];

struct nuddle_pq {              
  (*\textcolor{darkred}{nm\_oblv\_set}*) *base_pq;      
  int servers, groups, clnt_per_group; 
  int server_cnt, clients_cnt, group_cnt;   
  cache_line *requests[groups][clnt_per_group]; 
  cache_line *responses[groups];     
  lock *global_lock;                
};

struct client {
  cache_line *request, *response;   
  int clnt_pos;
};

struct server {        
  (*\textcolor{darkred}{nm\_oblv\_set}*) *base_pq;
  cache_line *my_clients[], *my_responses[];
  int my_groups, clnt_per_group;  
};
\end{lstlisting}
\caption{\arxiv{Helper} structures of \nuddle{}.}
\label{alg:structs}
\end{figure}

\textbf{Initialization \arxiv{Step}.} \arxiv{Figure~\ref{alg:init} describes the initialization functions of \nuddle.} {\fontfamily{lmtt}\selectfont \textit{initPQ()}} initializes (i) the \arxiv{underlying} data structure using the corresponding function of the \emph{base algorithm} (line 25), and (ii) the additional fields  of {\fontfamily{lmtt}\selectfont \textit{struct nuddle\_pq}}. \arxiv{For} this function, programmers \arxiv{need to} specify the number of server \arxiv{threads} and the maximum number of client \arxiv{threads} to properly allocate cache lines needed for communication among them. \arxiv{Programmers} also specify the size of the client \arxiv{thread} group \arxiv{(line 27), which is typically} 7 or 15, \arxiv{if the} cache line is 64 or 128 bytes, respectively. As explained in \arxiv{\ffwd}~\cite{ffwd}, \arxiv{assuming} 8-byte return values, a dedicated 64-byte \arxiv{(or 128-byte)} response cache line \arxiv{can be} shared between up to 7 \arxiv{(or 15)} client \arxiv{threads}, because it also has to include \arxiv{one} additional toggle bit for each client \arxiv{thread}. After initializing {\fontfamily{lmtt}\selectfont \textit{struct nuddle\_pq}}, each running thread calls either {\fontfamily{lmtt}\selectfont \textit{initClient()}} or {\fontfamily{lmtt}\selectfont \textit{initServer()}} depending on its role. \arxiv{Each thread} initializes its own \arxiv{helper} structure ({\fontfamily{lmtt}\selectfont \textit{struct client}} or {\fontfamily{lmtt}\selectfont \textit{struct server}}) with request and response cache lines of the corresponding shared arrays of {\fontfamily{lmtt}\selectfont \textit{struct nuddle\_pq}}. Server \arxiv{threads} undertake client \arxiv{thread} groups with a round robin fashion, such that the load associated with client \arxiv{threads} is balanced among them. In function {\fontfamily{lmtt}\selectfont \textit{initServer()}}, it is the programmer's responsibility to properly pin \arxiv{software server threads} to hardware \arxiv{contexts} (line 56), such that \arxiv{server threads} are located in the \emph{same} NUMA node, and \arxiv{the programmer} fully benefits from the \nuddle{} technique. Moreover, \arxiv{given that} client \arxiv{threads} of the \emph{same} \arxiv{client thread} group share the same response cache line, the programmer \arxiv{could} pin \arxiv{client threads of the \emph{same} client thread group to hardware contexts of} the \emph{same} NUMA node to \arxiv{minimize} cache coherence \arxiv{overheads}. Finally, \arxiv{since} the request and response arrays of {\fontfamily{lmtt}\selectfont \textit{struct nuddle\_pq}} are \emph{shared} between all threads, a global lock is used when \arxiv{updating} them to ensure mutual exclusion.

\begin{figure}[H]
\begin{lstlisting}[language=C]
struct nuddle_pq *initPQ(int servers, int max_clients) {
  struct nuddle_pq *pq = allocate_nuddle_pq();
  (*\textcolor{darkred}{\_\_base\_init(pq->base\_pq);}*)
  pq->servers = servers;
  pq->clnt_per_group = client_group(cache_line_size);
  pq->groups = (max_clients +
    pq->clnt_per_group-1) / pq->clnt_per_group;
  pq->server_cnt = 0;
  pq->client_cnt = 0;
  pq->group_cnt = 0;
  pq->requests = malloc(groups * clnt_per_group);
  pq->responses = malloc(groups);
  init_lock(pq->global_lock);
  return pq;
}

struct client *initClient(struct nuddle_pq *pq) {
  struct client *cl = allocate_client();
  acquire_lock(pq->global_lock);
  cl->request = &(pq->requests[group_cnt][clients_cnt]);
  cl->response = &(pq->responses[group_cnt]);
  cl->pos = pq->client_cnt;
  pq->client_cnt++;
  if (pq->client_cnt % pq->clnt_per_group == 0) {
    pq->clients_cnt = 0;
    pq->group_cnt++;
  }
  release_lock(pq->global_lock);
  return cl;
}

struct server *initServer(struct nuddle_pq *pq, int core)
{
  set_affinity(core);
  struct server *srv = allocate_server();
  srv->base_pq = pq->base_pq;
  srv->my_groups = 0;
  srv->clnt_per_group = pq->clnt_per_group;
  acquire_lock(pq->global_lock);
  int j = 0;
  for(i = 0; i < pq->groups; i++)
    if(i % pq->servers == pq->server_cnt) {
      srv->my_clients[j] = pq->requests[i][0..gr_clnt];
      srv->my_responses[j++] = pq->responses[i];
      srv->my_groups++;
     }
  pq->server_cnt++;
  release_lock(pq->global_lock);
  return srv;
}
\end{lstlisting}
\caption{\arxiv{Initialization functions} of \nuddle{}.}
\label{alg:init}
\end{figure}

\textbf{Main API.} Figure~\ref{alg:functions} \arxiv{shows the core functions of \nuddle, where} we omit the \arxiv{corresponding functions for} \delete{} \arxiv{operation}, since they are very similar \arxiv{to that of} \insrt{} operation. \arxiv{Both} \insrt{} and \delete{} \arxiv{operations of \nuddle{}} have similar API with the classic API of \arxiv{prior} state-of-the-art priority queue \arxiv{implementations~\cite{spraylist,lotan_shavit,sagonas,linden_jonsson}}. However, we separate the corresponding \arxiv{functions} for client \arxiv{threads} and server \arxiv{threads}. A client \arxiv{thread} writes its request to a dedicated request cache line (line 75) and then waits for the server \arxiv{thread}'s response. \arxiv{In contrast, a} server \arxiv{thread} directly executes operations in the data structure using the core \arxiv{functions} of the \emph{base algorithm} (line 82). \arxiv{Moreover, a server thread} can serve client \arxiv{threads} using the {\fontfamily{lmtt}\selectfont\textit{serve\_requests()}} function. A server \arxiv{thread} iterates \arxiv{over its own} client \arxiv{thread} groups and executes the requested operations in the data structure. \arxiv{The server thread} buffers individual return values \arxiv{for clients} to a local cache line ({\fontfamily{lmtt}\selectfont \textit{resp}} in lines 92 and 94) until \arxiv{it finishes processing all requests for the current client \arxiv{thread} group. Then, it} writes all responses to the \emph{shared} response cache line of that client \arxiv{thread} group \arxiv{(line 96)}, and proceeds to \arxiv{its} next client \arxiv{thread} group.

\begin{figure}[H]
\begin{lstlisting}[language=C]
int insert_client(struct client *cl,                                   int key, (*\textcolor{blue}{\textbf{\scriptsize{int64\_t}}}*) value)
{
  cl->request = write_req("insert", key, value);
  while (cl->response[cl->pos] == 0) ;
  return cl->response[cl->pos];
}

int insert_server(struct server *srv,                                  int key, (*\textcolor{blue}{\textbf{\scriptsize{int64\_t}}}*) value)
{
  return (*\textcolor{darkred}{\_\_base\_insrt}*)(srv->base_pq, key, value);
}

void serve_requests(struct server *srv) {
  for(i = 0; i < srv->mygroups; i++) {
    cache_line resp;
    for(j = 0; j < srv->clnt_per_group; j++) {
      key = srv->my_clients[i][j].key;
      value = srv->my_clients[i][j].value;
      if (srv->my_clients[i][j].op == "insert")
        resp[j] = (*\textcolor{darkred}{\_\_base\_insrt}*)(srv->base_pq, key, value);  
      else if (srv->my_clients[i][j].op == "deleteMin")
        resp[j] = (*\textcolor{darkred}{\_\_base\_delMin}*)(srv->base_pq);
    }
    srv->my_responses[i] = resp;
  }
}
\end{lstlisting}
\caption{Functions used by server \arxiv{threads} and client \arxiv{threads to perform operations using} \nuddle{}.}
\label{alg:functions}
\end{figure}

\begin{figure*}
    \centering\captionsetup[subfloat]{labelfont=bf}
  \begin{subfigure}[h]{0.48\textwidth}
    \centering
    \includegraphics[scale=0.154]{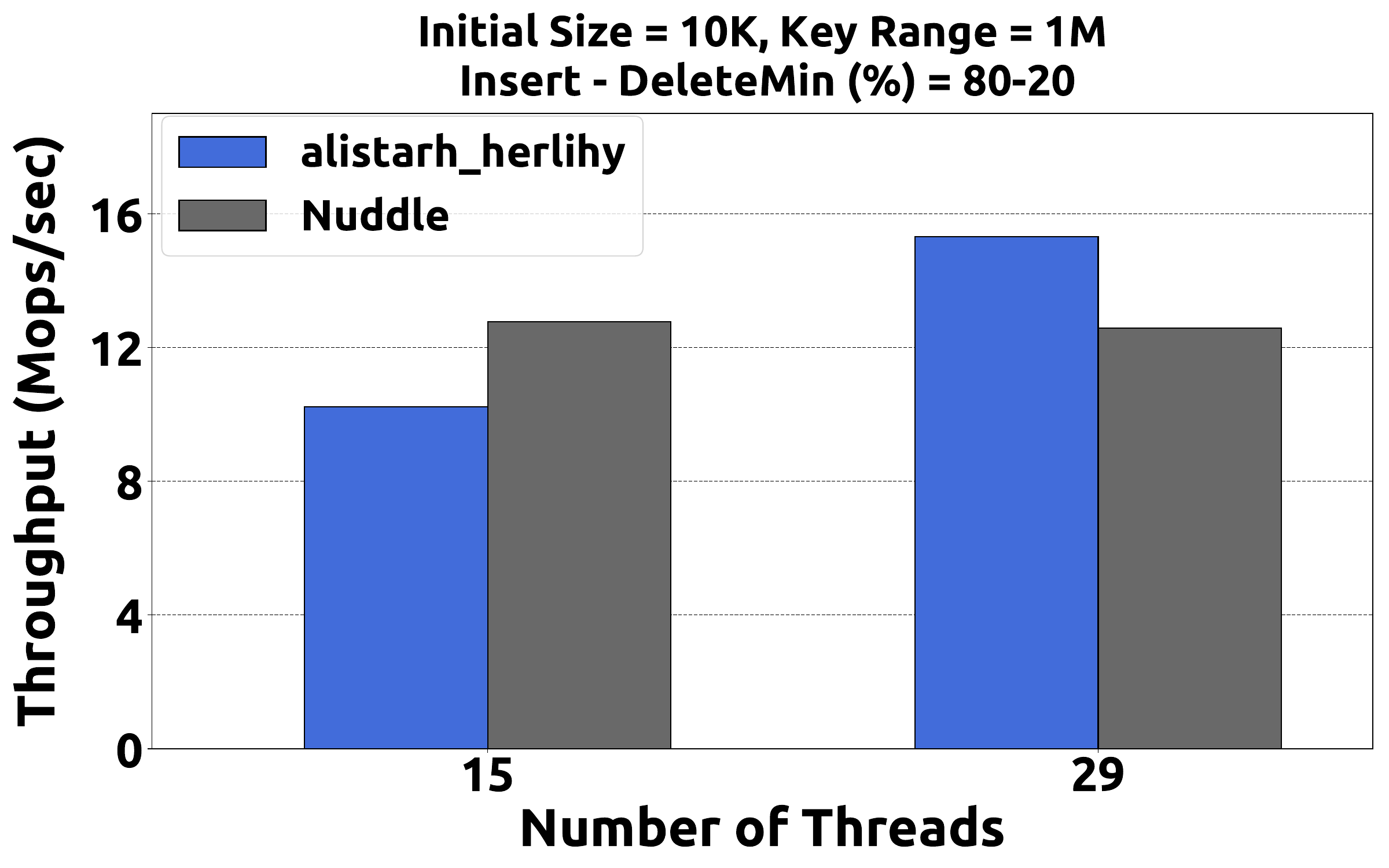}
    \vspace{-6pt}
    \caption{} 
 \end{subfigure}
  ~
  \begin{subfigure}[h]{0.48\textwidth}
   \centering
    \includegraphics[scale=0.154]{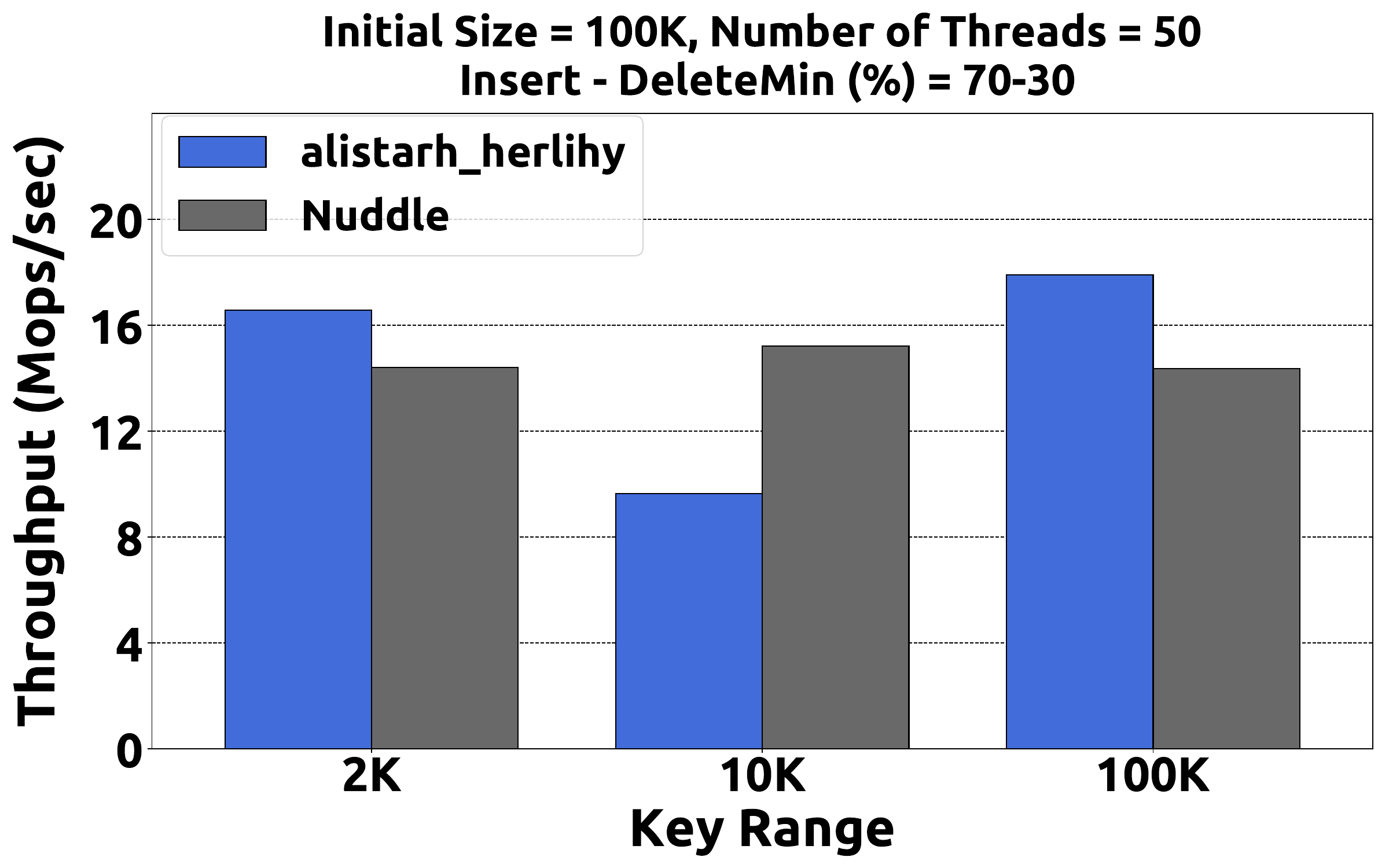}
    \vspace{-6pt}
    \caption{}
  \end{subfigure} 
  \caption{\arxiv{Throughput achieved by} \nuddle{} \arxiv{(using 8 server threads)} and its underlying \notnuma{} \arxiv{\emph{base algorithm}, \arxiv{i.e., \emph{alistarh\_herlihy}~\cite{spraylist,herlihy},} when we vary} (a) the number of threads that perform operations in the shared data structure, and (b) the key \arxiv{range of the workload.}}
  \label{fig:nuddle_spraylist}
  \vspace{0pt}
\end{figure*}

\section{\smartpq} \label{SmartPQbl}

We propose \arxiv{\smartpq, an adaptive concurrent priority queue which tunes itself by \emph{dynamically} switching between \notnuma{} and \numa{} \algomode{} modes, in order to perform best in \emph{all} contention workloads and at \emph{any} point in time, even when contention varies over time.}

Designing an adaptive priority queue \arxiv{involves addressing two major challenges: (i)} how to switch \arxiv{from one \algomode{} mode to the other} with \emph{low overhead}, and \arxiv{(ii) \emph{when} to switch from one \algomode{} mode to the other. 

To address the first challenge, we exploit the fact that the actual underlying implementation of \nuddle{} is a \emph{concurrent} \notnuma{} implementation. We select \nuddle{}, as the \numa{} \algomode{} mode of \smartpq{}, and its underlying \emph{base algorithm}, as the \notnuma{} \algomode{} mode of \smartpq. Threads can perform operations in the data structure using either \nuddle{} or its underlying \emph{base algorithm}, with \emph{no actual change} in the way data is accessed. As a result, \smartpq{} can switch between the two \algomode{} modes \emph{without} needing a synchronization point between transitions, and without violating correctness.

To address the second challenge, we design a simple decision tree classifier (Section~\ref{sec:classifier}), and train it to select the best-performing \algomode{} mode between \nuddle{}, as the \numa{} \algomode{} mode of \smartpq, and its underlying \emph{base algorithm}, as the \notnuma{} mode of \smartpq. Finally, we add a \emph{lightweight} decision making mechanism in \smartpq{} (Section~\ref{sec:smartpq_details}) to dynamically tune itself over time between the two \algomode{} modes. We describe more details in next sections.
}

\subsection{Selecting \arxiv{t}he \arxiv{A}lgorithmic Mode}

\subsubsection{The Need for \arxiv{a} Machine Learning \arxiv{Approach}}

Selecting \arxiv{the best-performing algorithmic mode can be solved in various ways. For instance, one could take an empirical \arxiv{exhaustive} approach: measure the throughput achieved by the two algorithmic modes for all various contention scenarios on the target NUMA system, and then use the \algomode{} mode that achieves the highest throughput on future runs of the same contention workload on the target NUMA system. Even though this is the most accurate method, it (i) incurs \emph{substantial} overhead and effort to sweep over \emph{all} various contention workloads, and (ii) would need a large amount of memory to store the best-performing \algomode{} mode for all various scenarios. Furthermore, it is not trivial to construct a statistical model to predict the best-performing \algomode{} mode, since the performance of an algorithm is also affected by the characteristics of underlying computing platform. Figure~\ref{fig:nuddle_spraylist} summarizes these observations by comparing \nuddle{}  with its underlying \emph{base algorithm} in a 4-node NUMA system. For the \emph{base algorithm}, we use \emph{alistarh\_herlihy} priority queue~\cite{spraylist,herlihy}, since this is the \notnuma{} implementation that achieves the highest performance, according to our evaluation (Section~\ref{sec:experimental}). 
}

Figure~\ref{fig:nuddle_spraylist}a \arxiv{demonstrates that the best-performing \algomode{} mode depends on multiple parameters, such as the number of threads that perform operations in the shared data structure. \arxiv{We find that the \algomode{} also depends on} the size of the data structure, and the operation workload, i.e., the \arxiv{percentage} of \insrt/\delete{} operations. Specifically,} when the number of threads increases, we may expect that the performance of \arxiv{the \notnuma{}} \emph{alistarh\_herlihy} degrades \arxiv{due to higher contention}. \arxiv{In contrast, with 80\% \insrt{} operations} when \arxiv{increasing the number of threads to} 29, \emph{alistarh\_herlihy} outperforms \nuddle. \arxiv{This is because} the size of the priority queue and the range \arxiv{of keys used in the workload} are relatively large, \arxiv{while the percentage} of \delete{} operations is low. \arxiv{In this scenario}, threads \arxiv{may not compete for the same elements, working on different parts of the data structure,} and thus, \arxiv{the \notnuma{}} \emph{alistarh\_herlihy} achieves higher throughput \arxiv{compared to the \numa{} \nuddle.}

Figure~\ref{fig:nuddle_spraylist}b \arxiv{demonstrates that the best-performing \algomode{} mode cannot be straightforwardly predicted, and also depends on the characteristics of the workload and of the underlying hardware.} In \insrt-dominated workloads, as the key range increases, \arxiv{threads may update different parts of the shared data structure.} We might, thus, expect that after a certain point \arxiv{of increasing the key range, the \notnuma{}} \emph{alistarh\_herlihy} will always outperform \nuddle, \arxiv{since} the contention decreases. However, \arxiv{we note that, even though} the performance of \nuddle{} remains constant, as expected, the performance of \emph{alistarh\_herlihy} \arxiv{highly} varies as the key range \arxiv{increases} due to the hyperthreading effect. When using more than 32 threads, hyperthreading is enabled in our NUMA system (Section~\ref{sec:experimental}). The hyperthreading pair of threads shares the L1 and L2 caches\arxiv{, and thus, these threads} may either thrash or benefit from each other depending on \arxiv{the characteristics of L1 and L2 caches (e.g., size, eviction policy), and} the elements accessed in each operation.


Considering \arxiv{the aforementioned} non-straightforward behavior, we resort to \arxiv{a} machine learning \arxiv{approach} as the basis of \arxiv{our} prediction mechanism.

\subsubsection{Decision Tree Classifier}\label{sec:classifier}
We formulate the selection of the \algomode{} mode as a classification problem, and leverage supervised learning techniques to train a \arxiv{simple} classifier \arxiv{to predict} the \arxiv{best-performing} \algomode{} mode for each \arxiv{contention} workload. For our classifier, we \arxiv{select} decision trees, since they are commonly used \arxiv{in classification models for multithreaded workloads}~\cite{athena,Doddipalli2012Ensemble,Polat2009ANovel,Meng2019APattern,Dhulipala2020APattern,Sloan2012Algorithmic,Sedaghati2015Automatic,Benatia2016SparseMF}, \arxiv{and incur} low training and inference \arxiv{overhead}. Moreover, they are easy to interpret and thus, be incorporated to \arxiv{our} proposed priority queue (Section~\ref{sec:smartpq_details}). We generate the decision tree classifier using the scikit-learn machine learning toolkit~\cite{scikit}.

\textbf{\textit{1) Class Definition:}} 
\arxiv{We define the following classes:} (a) the \textbf{\notnuma} class that stands for the \notnuma{} \algomode{} mode, (b) the \textbf{\numa{}} class that stands for the \numa{} \algomode{} mode, and (c) the \textbf{\textit{neutral}} class that stands for a tie, meaning that \arxiv{either a \numa{} or a \notnuma{} implementation can be selected, since they achieve similar performance. We include a neutral class for two reasons:} (i) when using \emph{only one} socket of \arxiv{a NUMA system}, \numa{} implementations deliver \arxiv{similar throughput with \notnuma{} implementations, and (ii) in an adaptive data structure, which \emph{dynamically} switches between the two \algomode{} modes,} we want to configure a transition from one \arxiv{\algomode{} mode} to another to occur when the difference in their throughput is \arxiv{relatively high, i.e.,} greater than a certain threshold. Otherwise, \arxiv{the adaptive data structure} might continuously oscillate between the two modes, without delivering significant performance improvements or even causing performance degradation.

\textbf{\textit{2) \arxiv{Extracted Features}}}: \arxiv{Table~\ref{table:features} \arxiv{explains} the \arxiv{four} features \arxiv{of the contention workload which are} used in our classifier \arxiv{targeting} priority queues. We assume that the contention workload is known a priori, and thus, we can easily extract the features needed for classification. Section~\ref{sec:discussion} discusses how to on-the-fly extract these features.}

\begin{table}[H]
\centering
\resizebox{\linewidth}{!}{
\begin{tabular}{l l} 
 \toprule
 Feature & Definition  \\ [0.5ex] \midrule \midrule
 \multirow{2}{*}{\shortstack[l]{\#Threads}} & The number of \arxiv{active} threads \\
  & \arxiv{that perform operations in the data structure} \\
 Size & The \arxiv{current} size of the priority queue\\
Key\_range & The range of keys used in \arxiv{the workload} \\
 \multirow{1}{*}{\shortstack[l]{\% \insrt /\delete}} & The \arxiv{percentage} of \insrt /\delete{} operations \\
 \bottomrule
\end{tabular}
}
\caption{The features \arxiv{of the contention workload which are} used for classification.}
\label{table:features}
\end{table}

\textbf{\textit{3) Generation of Training Data:}}
To train our classifier, we develop microbenchmarks, in which threads repeatedly execute random operations on the \arxiv{priority queue} for 5 seconds. We \arxiv{select} \nuddle{}, as the \numa{} \arxiv{implementation}, and \emph{alistarh\_herlihy}, as \arxiv{its underlying} \notnuma{} \arxiv{implementation, since this is the best-performing \notnuma{} priority queue (Section~\ref{sec:experimental}). We use} a variety of values for the features \arxiv{needed for classification} (Table~\ref{table:features}). Our training data set consists of 5525 different \arxiv{contention} workloads. Finally, we pin software threads to hardware \arxiv{contexts} \arxiv{of the evaluated NUMA system} in a round-robin fashion, and thus, the classifier is trained \arxiv{with} this thread placement. \arxiv{We leave the exploration of the thread placement policy for future work.}

\textbf{\textit{4) Labeling of Training Data:}}
Regarding \arxiv{the labeling of our training} data set, we set the threshold for tie between the two \algomode{} modes to \arxiv{an empirical value of} 1.5 Million operations per second. When the difference in throughput \arxiv{between the two \algomode{} modes} is less than this threshold, the \textit{neutral} class is selected as label. Otherwise, we select the class that corresponds to the \algomode{} mode that \arxiv{achieves} the \arxiv{highest} throughput.

The final decision tree classifier has \arxiv{only} 180 nodes, half of which are leaves. It has a very low depth of 8, \arxiv{that} is the length of the longest path in the tree, and thus, a \emph{very low} traversal cost (2-4 ms in \arxiv{our evaluated NUMA system).}

\subsection{Implementation Details} \label{sec:smartpq_details}

\begin{figure}
\hspace{8pt}
\begin{subfigure}{1\linewidth}
\begin{lstlisting}[language=C]
struct smartpq {              
  (*\textcolor{darkred}{nm\_oblv\_set}*) *base_pq;      
  int servers, groups, clnt_per_group; 
  int server_cnt, clients_cnt, group_cnt;   
  cache_line *requests[groups][clnt_per_group]; 
  cache_line *responses[groups];     
  lock *global_lock;    
  (*\textcolor{mygreen}{int *algo;}*) (*\textcolor{mygreen}{// 1: NUMA-oblivious (default), \\  2: NUMA-aware}*)
};

struct client {
  (*\textcolor{darkred}{nm\_oblv\_set}*) (*\textcolor{mygreen}{*base\_pq;}*)
  (*\textcolor{mygreen}{int *algo;}*)  
  cache_line *request, *response;   
  int clnt_pos;             
};

struct client *initClient(struct smartpq *pq) { 
  ... lines 40-49 of Fig. 5 ...
  (*\textcolor{mygreen}{cl->base\_pq = pq->base\_pq;}*)
  (*\textcolor{mygreen}{cl->algo = \&(pq->algo);}*)
  release_lock(pq->global_lock);  
  return cl; 
}

int insert_client(struct client *cl,                            int key, float value) {
  (*\textcolor{mygreen}{if(*(cl->algo) == 1) \{ }*)       
    return (*\textcolor{darkred}{\_\_base\_insert(cl->base\_pq,key,value);}*)
  (*\textcolor{mygreen}{\} else \{ // *(cl->algo) == 2}*)
    ... lines 75-77 of Fig. 6 ...
  (*\textcolor{mygreen}{\}}*)                         
}   

void serve_requests(struct server *srv) {
  (*\textcolor{mygreen}{if(*(srv->algo) == 2)\{}*) 
    for(i = 0; i < srv->mygroups; i++) {
      cache_line resp;
      for(j = 0; j < srv->clnt_per_group; j++) {
        key = srv->my_clients[i][j].key;
        value = srv->my_clients[i][j].value;
        if (srv->my_clients[i][j].op == "insert")
          resp[j] = (*\textcolor{darkred}{\_\_base\_insrt}*)(srv->base_pq, key, value);  
        else if (srv->my_clients[i][j].op == "deleteMin")
          resp[j] = (*\textcolor{darkred}{\_\_base\_delMin}*)(srv->base_pq);
      }
      srv->my_responses[i] = resp;
    }
  (*\textcolor{mygreen}{\} else}*)
    (*\textcolor{mygreen}{return;}*)
}

void decisionTree(struct server / (* \\ *)           struct client *str, int nthreads,  (* \\ *)             int size, int key_range, (* \\ *)            double insert\_deleteMin) {
  (*\textcolor{mygreen}{int algo = 0;}*)
  (*\textcolor{mygreen}{... code for decision tree classifier ...}*)
  (*\textcolor{mygreen}{if (algo != 0) //  0: neutral}*)  
    (*\textcolor{mygreen}{*(str->algo) = algo;}*)
}
\end{lstlisting}
\end{subfigure}
\caption{The modified code of \nuddle{} \arxiv{adding the decision making mechanism} to implement \smartpq{}.}
\label{alg:smartpq}
\end{figure}

Figure~\ref{alg:smartpq} presents the modified code of \nuddle{} \arxiv{adding the decision making mechanism (using green color)} to implement \smartpq.
We extend the main structure of \nuddle{}, renamed to {\fontfamily{lmtt}\selectfont \textit{struct smartpq}}, by adding an additional field, called {\fontfamily{lmtt}\selectfont \textit{algo}}, \arxiv{to keep} track the current \algomode{} mode, \arxiv{(either \notnuma{} or \numa{})}. Similarly, {\fontfamily{lmtt}\selectfont \textit{struct client}} and {\fontfamily{lmtt}\selectfont \textit{struct server}} \arxiv{structures} are extended with an additional {\fontfamily{lmtt}\selectfont \textit{algo}} field \arxiv{(e.g., line 111)}, \arxiv{which} is a pointer to the {\fontfamily{lmtt}\selectfont \textit{algo}} \arxiv{field of {\fontfamily{lmtt}\selectfont \textit{struct smartpq}}}. \arxiv{Each active thread initializes this pointer either} in {\fontfamily{lmtt}\selectfont \textit{initClient()}} \arxiv{or} {\fontfamily{lmtt}\selectfont \textit{initServer()}} depending on its role (e.g., line 119). \arxiv{This} way, all threads share the same \algomode{} mode at \emph{any} point in time. \arxiv{In {\fontfamily{lmtt}\selectfont \textit{struct client}}, we also add a pointer to the shared data structure (line 110), which is used by client threads to \emph{directly} perform operations in the data structure in case of \notnuma{} mode. Specifically,} we modify the core \arxiv{functions} of client \arxiv{threads, i.e., {\fontfamily{lmtt}\selectfont \textit{insert\_client()}} and {\fontfamily{lmtt}\selectfont \textit{deleteMin\_client()}}}, such that \arxiv{client threads} either directly execute their \arxiv{operations} in the data structure (e.g., line 126), or delegate them to server \arxiv{threads} (e.g., line 127-128), with respect to the current \algomode{} mode. \arxiv{In contrast}, the core \arxiv{functions} of server \arxiv{threads} do not need any modification. Finally, we wrap the code of {\fontfamily{lmtt}\selectfont \textit{serve\_requests}} function, i.e., the lines \arxiv{86-97} of Figure~\ref{alg:functions}, with an if/else statement on the {\fontfamily{lmtt}\selectfont \textit{algo}} field \arxiv{(lines 133, 146 in Fig.~\ref{alg:smartpq})}, such that server \arxiv{threads} poll at client \arxiv{threads}' requests only in \numa{} mode. \arxiv{In} \notnuma{} mode, {\fontfamily{lmtt}\selectfont \textit{serve\_requests}} function returns without doing \arxiv{nothing}. \arxiv{This} way, programmers do not need to take care of calls on this function in their code, when the \notnuma{} mode is selected.

\arxiv{The} {\fontfamily{lmtt}\selectfont \textit{decisionTree()}} \arxiv{function} describes the \arxiv{interface with our proposed decision tree} classifier, where the \arxiv{input} arguments are associated with its features. In frequent time lapses, one or more threads \arxiv{may} call this function to check if a transition to another \arxiv{\algomode{}} mode is needed. If this is the case, the {\fontfamily{lmtt}\selectfont \textit{algo}} field of {\fontfamily{lmtt}\selectfont \textit{struct smartpq}} is updated \arxiv{(line 154 in Fig.~\ref{alg:smartpq})}, \arxiv{and} \smartpq{} switches \algomode{} mode, \arxiv{i.e.,} all \arxiv{active} threads start executing their operations using the new \algomode{} mode. If the classifier predicts the neutral class \arxiv{(line 153)}, the {\fontfamily{lmtt}\selectfont \textit{algo}} field is not updated, and \arxiv{thus} \smartpq{} remains at the \arxiv{currently} selected \arxiv{\algomode{}} mode.

\section{Experimental Evaluation}
\label{sec:experimental}

In our experimental evaluation, we use a 4-socket Intel Sandy Bridge-EP server equipped with 8-core Intel Xeon CPU E5-4620 processors providing a total of 32 physical cores and 64 hardware contexts. The processor runs at 2.2GHz and each physical core has its own L1 and L2 cache of sizes 64KB and 256KB, respectively. A 16MB L3 cache is shared by all cores in a NUMA socket and the RAM is 256GB. We use GCC 4.9.2 with -O3 optimization flag enabled to compile \arxiv{all} implementations.


Our evaluation includes the following concurrent priority queue \arxiv{implementations}:
\begin{compactitem}[$\textbf{--}$]
\item \textit{alistarh\_fraser}~\cite{fraser,spraylist}: A \notnuma, relaxed priority queue~\cite{spraylist} based on Fraser's skip-list~\cite{fraser} available at \arxiv{ASCYLIB library}~\cite{ascylib}.
\item \textit{alistarh\_herlihy}~\cite{herlihy,spraylist}: A \notnuma, relaxed priority queue~\cite{spraylist} based on Herlihy's skip-list~\cite{herlihy} available at \arxiv{ASCYLIB library}~\cite{ascylib}.
\item \textit{lotan\_shavit}~\cite{lotan_shavit}: A \notnuma{} priority queue available at \arxiv{ASCYLIB library}~\cite{ascylib}.
\item \ffwd~\cite{ffwd}: A \numa{} \arxiv{priority queue} based on \arxiv{the delegation technique~\cite{Calciu2013Message,Klaftenegger2014Delegation,Lozi2012Remote,Petrovic2015Performance,Suleman2009Accelerating}, which includes \emph{only}} one server \arxiv{thread to perform operations on behalf of \emph{all} client threads}.
\item \nuddle: \arxiv{Our} proposed \numa{} priority queue, which uses \emph{alistarh\_herlihy} as the underlying \emph{base algorithm}.
\item \smartpq: \arxiv{Our} proposed adaptive priority queue, which uses \nuddle{} as the \numa{} mode, and \emph{alistarh\_herlihy} as the \notnuma{} \emph{base algorithm}.
\end{compactitem}

We evaluate the concurrent priority queue \arxiv{implementations} in the following way:
\begin{compactitem}[$\textbf{--}$]
\item Each execution lasts 5 seconds, during which each thread performs randomly chosen operations. We also tried longer durations and got similar results.
\item \arxiv{Between consecutive operations in the data structure each thread executes} a delay loop of 25 pause instructions. This delay is intentionally added \arxiv{in} our benchmarks to better simulate a real-life application, where operations in the data structure are intermingled with \arxiv{other} instructions \arxiv{in} the application.
\item At the beginning of each run, the priority queue is initialized with elements the number of which is described at each figure.
\item Each software thread is pinned to a hardware \arxiv{context}. Hyperthreading is enabled when using more than 32 \arxiv{software} threads. When exceeding the number of available hardware contexts \arxiv{of the system}, we oversubscribe \arxiv{software threads to hardware contexts.}
\item \arxiv{We pin the} first 8 threads to the first \arxiv{NUMA node,} and \arxiv{consecutive client thread groups of 7 client threads each, to NUMA nodes in a round-robin fashion.} 
\item In \notnuma{} implementations, any allocation needed in the operation is executed on demand, and memory affinity is determined by the first touch policy.
\item In \numa{} implementations, since our \arxiv{NUMA system} has 64-byte cache lines, the response cache line is shared \arxiv{between up to} 7 client \arxiv{threads, using} 8-byte return values.
\item In \nuddle, the first 8 threads represent server \arxiv{threads. Server threads repeatedly execute the}  {\fontfamily{lmtt}\selectfont \textit{serve\_requests}} \arxiv{function, and then} a randomly chosen \arxiv{operation} until time is up.
\item We have disabled the automatic Linux Balancing \cite{numa-balancing} to get consistent and stable results.
\item All reported results are the average of 10 independent executions with no significant variance.
\end{compactitem}

\begin{figure*}[ht]
\centering
\includegraphics[scale=0.24]{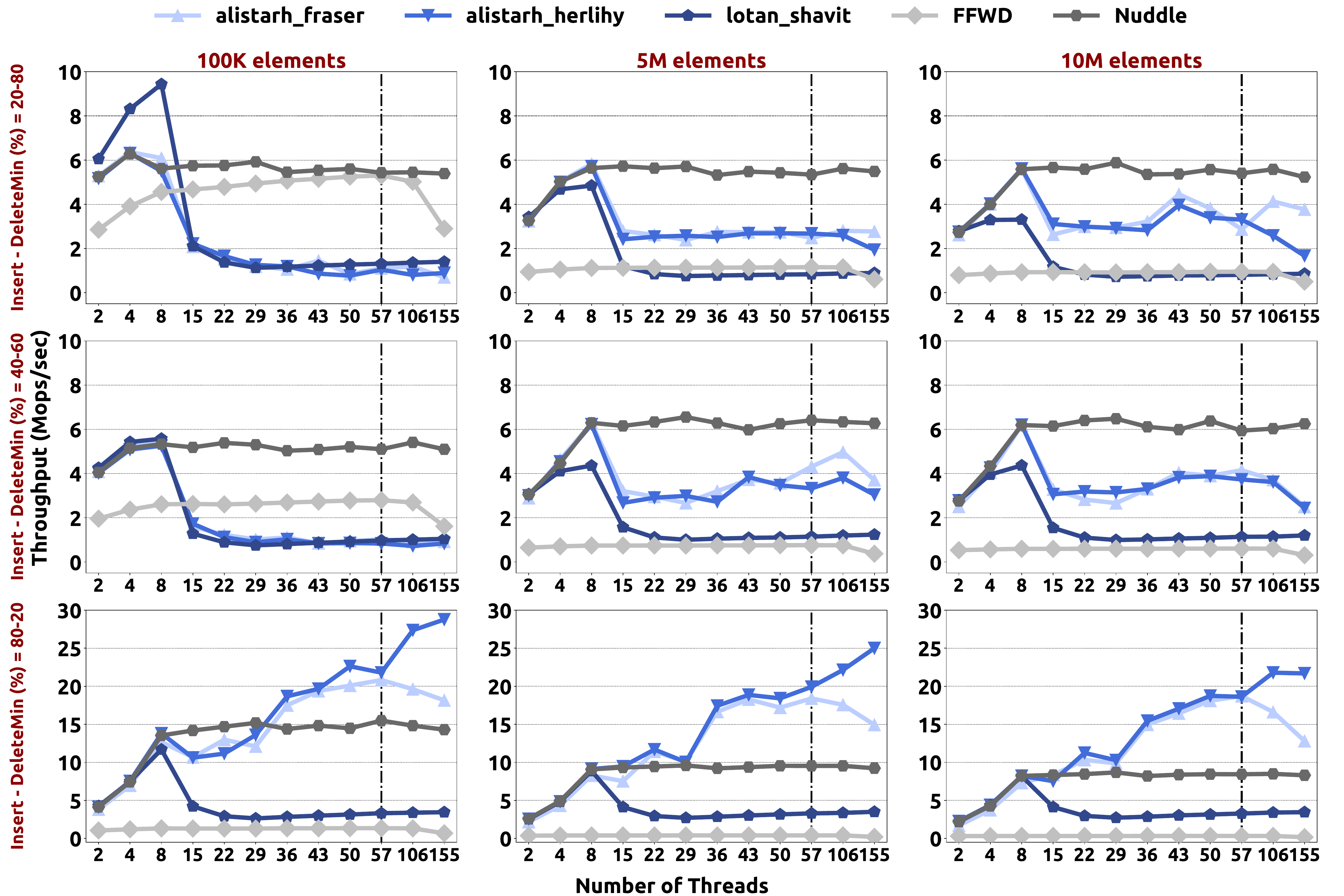}
\caption{Throughput of concurrent priority queue implementations. The columns \arxiv{show} different \arxiv{priority queue sizes using the key range of double the elements of each size. The} rows \arxiv{show} different operation workloads. The vertical line in each plot shows the point after which we oversubscribe software threads to hardware \arxiv{contexts}.}
\label{fig:simple_bench}
\end{figure*}

\subsection{Throughput of \nuddle}\label{sec:simbple_bench}
Figure~\ref{fig:simple_bench} presents the throughput achieved by concurrent priority queue \arxiv{implementations} for various sizes and operation workloads. \arxiv{\numa{} priority queue implementations, i.e., \ffwd{} and \nuddle,} achieve high throughput in \delete-dominated workloads: \nuddle{} performs best in \emph{all} \delete-dominated workloads, while \ffwd{} outperforms \notnuma{} implementations \arxiv{in the small-sized} priority queues (\arxiv{e.g.,} 100K elements). \arxiv{In large-sized priority queues, \insrt{} operations have a larger impact on the total execution time (due to a longer traversal), and thus \nuddle{} and \notnuma{} implementations perform better than \ffwd, since they provide \emph{higher} thread-level parallelism. Note that \ffwd{} has \emph{single-threaded} performance, since at any point in time only \emph{one} (server) thread performs operations in the data structure. Moreover, as it is expected, the performance of both \ffwd{} and \nuddle{} saturates at the number of server threads used (e.g., 8 server threads for \nuddle) to perform operations in the data structure. Finally, we note that the communication between server and client threads used in \numa{} implementations} has negligible overhead; when \arxiv{the number of} client \arxiv{threads increases}, even though the communication traffic \arxiv{over the interconnect} increases, there is \emph{no} performance drop. \arxiv{Overall, we} conclude that \nuddle{} achieves the highest throughput in \emph{all} \delete-dominated workloads, \arxiv{and is the most efficient \numa{} approach, since it provides high thread-level parallelism.}

On the other hand, \notnuma{} \arxiv{implementations incur high performance degradation in \emph{high-contention} scenarios, such as \delete-dominated workloads, \arxiv{when using more than one NUMA node (i.e., after 8 threads)}.  As already discussed in prior works~\cite{David2013Everything,Boyd2010AnAnalysis,Lozi2012Remote,Zhang2016Scalable,Molka2009Memory,giannoula2021SynCron}, the non-uniformity in memory accesses and cache line invalidation traffic significantly affects performance in high-contention scenarios.} In \insrt-dominated workloads, \arxiv{which incur lower contention, even though} \emph{lotan\_shavit} priority queue \arxiv{still incurs performance degradation} when using more than one \arxiv{NUMA nodes of the system}, the \emph{relaxed} \notnuma{} \arxiv{implementations, i.e., \emph{alistarh\_fraser} and \emph{alistarh\_herlihy} priority queues,} achieve high scalability. \arxiv{This is because relaxed priority queues decrease both (i) the contention among threads, and (ii) the cache line invalidation traffic:} \arxiv{the} \delete{} operation returns (with a high probability) an element among the \emph{first few} (high-priority) elements of the queue, and thus, threads do not frequently compete for the same elements. \arxiv{Finally, we observe that \emph{alistarh\_herlihy} priority queue achieves higher performance benefits over \emph{alistarh\_fraser} priority queue, when we oversubscribe software threads to the available hardware contexts of our system. Overall, we find that in \insrt-dominated workloads, the \emph{relaxed} \notnuma{} implementations significantly outperform the \numa{} ones.}

To \arxiv{sum up, we conclude} that there is no one-size-fits-all solution, since none of the priority queues \arxiv{performs best across \emph{all} contention} workloads. \nuddle{} \arxiv{achieves the highest throughput in high contention scenarios, while \emph{alistarh\_herlihy} performs best in low and medium contention scenarios.} It is thus desirable to \arxiv{design a new approach for a concurrent priority queue to perform} best \arxiv{under \emph{all} various contention scenarios.}

\subsection{Throughput of \smartpq}

\subsubsection{Classifier Accuracy}\label{sec:accuracy}
We evaluate \arxiv{the efficiency} of our proposed classifier \arxiv{(Section~\ref{sec:classifier}) using two metrics: (i) accuracy, and (ii) misprediction cost. First, we define the accuracy of the classifier as} the percentage of \emph{correct} predictions, \arxiv{where a prediction is considered correct,} if the classifier predicts the \algomode{} mode (either the \numa{} \nuddle{} or the \notnuma{} \emph{alistarh\_herlihy}) \arxiv{that achieves} the \arxiv{best performance between the two.} We use a test set of 10780 different \arxiv{contention} workloads, \arxiv{where} we randomly select the values of the features \arxiv{in each workload.} In the above test set, our classifier has 87.9\% accuracy, \arxiv{i.e.,} it mispredicts 1300 times in 10780 different \arxiv{contention} workloads. \arxiv{Second, we define the misprediction cost as the performance difference between the correct \arxiv{(best-performing) \algomode{} mode} and the wrong predicted mode normalized to the performance of the wrong predicted mode.} \arxiv{Specifically}, \arxiv{assuming} the throughput of the wrong predicted and correct \arxiv{(best-performing)} \algomode{} mode is $Y$ and $X$ respectively, the misprediction cost is \arxiv{defined as} $((X-Y)/Y) * 100\%$. In 1300 mispredicted workloads, the geometric mean of misprediction cost \arxiv{for our classifier} is 30.2\%. We conclude that \arxiv{the proposed} classifier has \emph{high} accuracy, and in case of misprediction, \arxiv{incurs} low performance \arxiv{degradation}.

\subsubsection{Varying the Contention Workload}

\begin{figure*}[h!]
\captionsetup[subfigure]{justification=centering}
    \begin{subfigure}[t]{0.33\textwidth}\centering
        \includegraphics[scale=0.3]{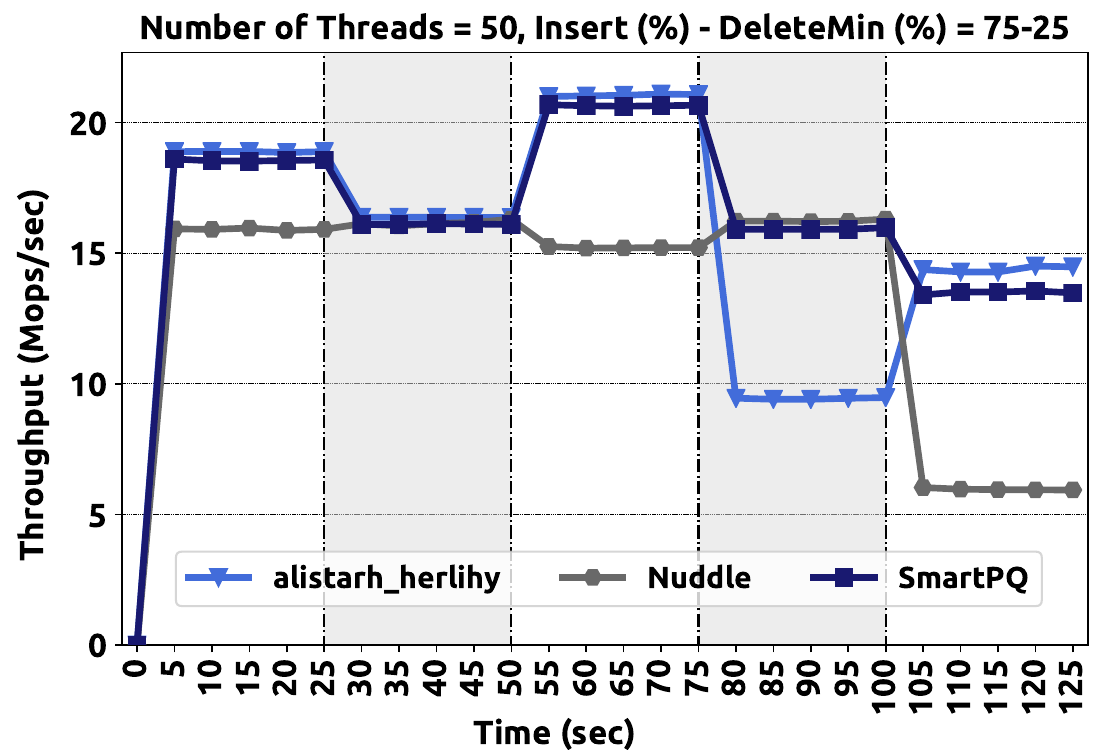}
        \vspace{-5pt}
        \caption{}
	\label{fig:dynamic_range}
    \end{subfigure}%
    ~
    \begin{subfigure}[t]{0.33\textwidth}\centering
        \includegraphics[scale=0.3]{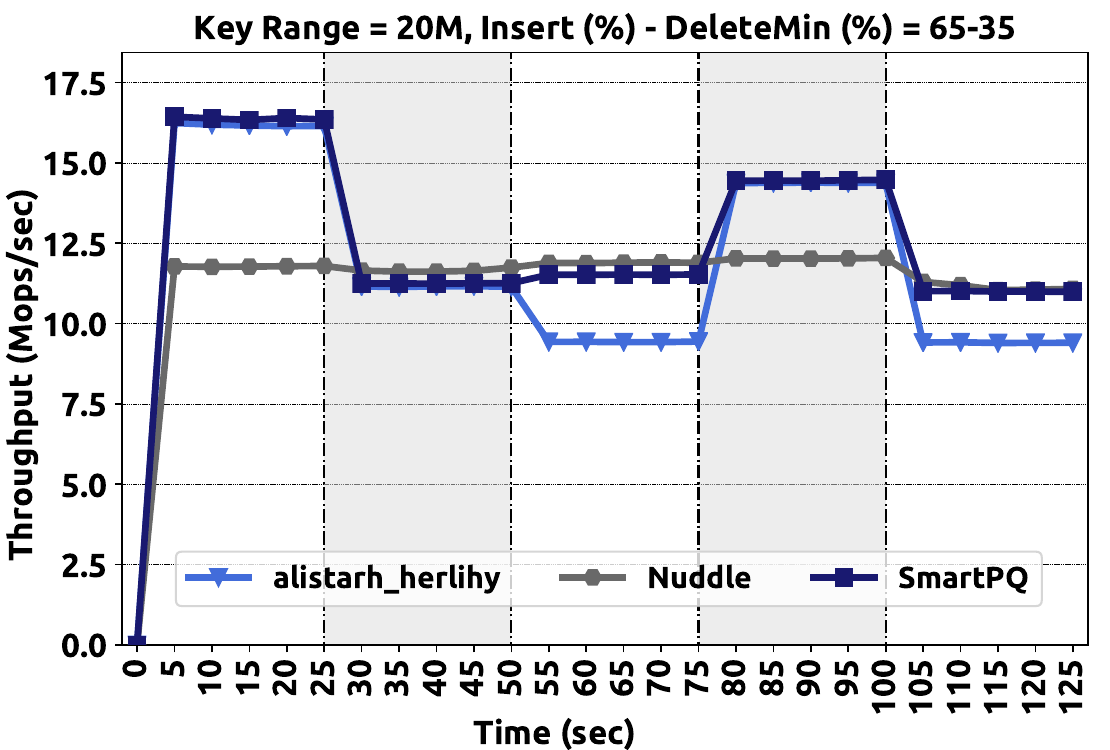}
        \vspace{-5pt}
        \caption{}
	\label{fig:dynamic_threads}
    \end{subfigure}
    ~ 
    \begin{subfigure}[t]{0.33\textwidth}\centering
        \includegraphics[scale=0.3]{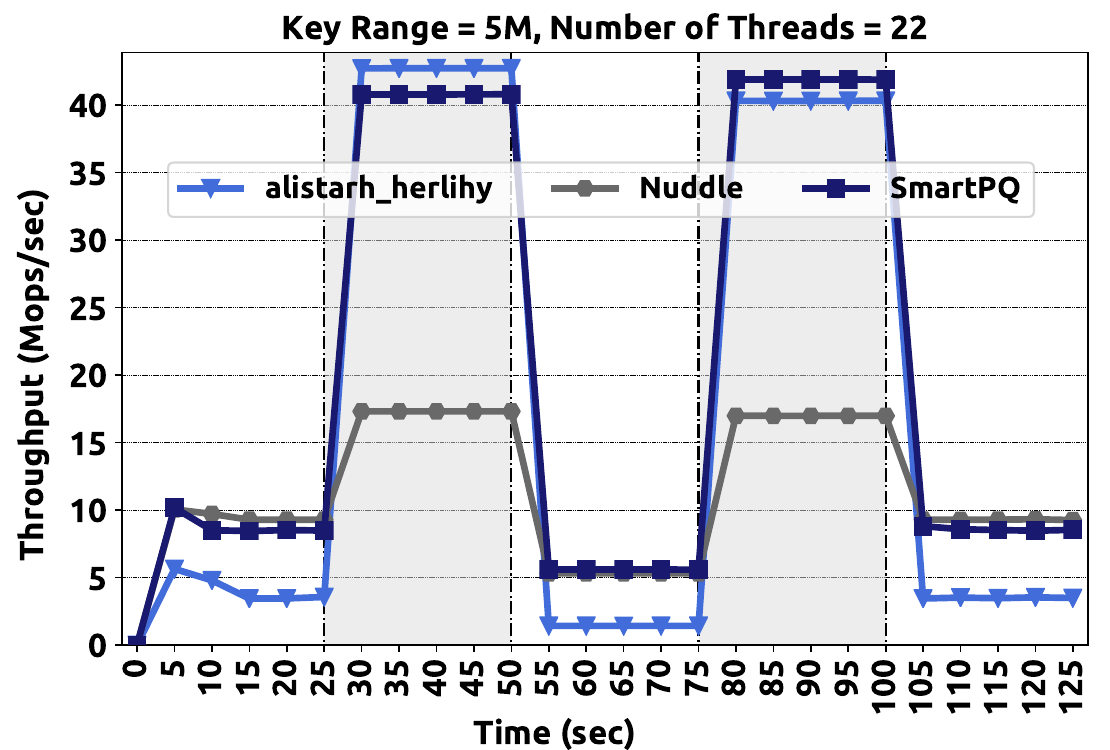}
        \vspace{-5pt}
        \caption{}
	\label{fig:dynamic_operation}
    \end{subfigure}
    \caption{Throughput achieved by \smartpq{}, \nuddle{} and \arxiv{its underlying} \emph{base algorithm} (\emph{alistarh\_herlihy}), in synthetic benchmarks, \arxiv{in which we vary} a) the key range, b) the number of threads that perform operations in the data structure, and c) the percentage of \insrt/\delete{} operations in the workload.}
    \label{fig:dynamic_bench}
\end{figure*}

\begin{figure*}[ht]
\centering
\includegraphics[scale=0.32]{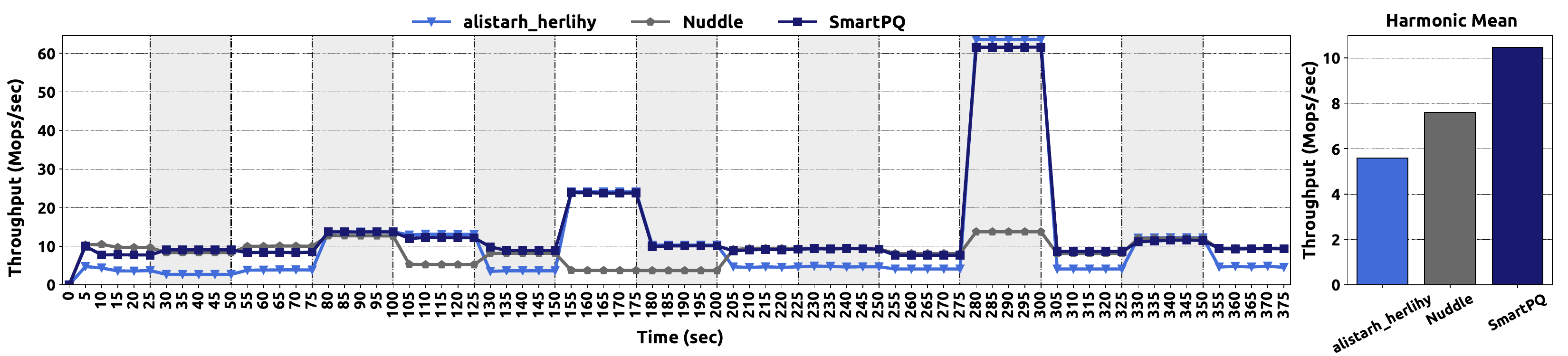}
\caption{Throughput \arxiv{achieved by} \smartpq{}, \nuddle{} and \arxiv{its underlying} \emph{base algorithm} (\emph{alistarh\_herlihy}), in synthetic benchmarks, \arxiv{in which we vary multiple features in the contention workload}.}
\label{fig:dynamic_total}
\end{figure*}

We present the performance benefit of \smartpq{} in \arxiv{synthetic} benchmarks, in which we vary the \arxiv{contention} workload \arxiv{over time}, and compare it with \nuddle{} and \arxiv{its underlying \emph{base algorithm}, i.e.,} \emph{alistarh\_herlihy} priority queue. In all benchmarks, we change the \arxiv{contention} workload every 25 seconds. \arxiv{In \smartpq, we set one dedicated sever thread to call the decision tree classifier \emph{every} second, in order to check if a transition to another \algomode{} mode is needed. Figure~\ref{fig:dynamic_bench} and Figure~\ref{fig:dynamic_total} show the throughput achieved by all three schemes, when we vary one and multiple features in the contention workload, respectively. Table~\ref{tab:features_dynamic_bench} and Table~\ref{tab:features_dynamic_total} show the features of the workload as they vary during the execution for the benchmarks evaluated in Figure~\ref{fig:dynamic_bench} and Figure~\ref{fig:dynamic_total}, respectively. Note that the current size of the priority queue changes during the execution due to successful \insrt{} and \delete{} operations. 
}

We make \arxiv{three observations. First, as already shown in Section~\ref{sec:simbple_bench},} there is no one-size-fits-all solution, since neither \nuddle{} nor \emph{alistarh\_herlihy} \arxiv{performs best across all various contention workloads. For instance, in} Figure~\ref{fig:dynamic_threads}, even though the performance of \nuddle{} \arxiv{remains} constant, it outperforms \emph{alistarh\_herlihy}, when \arxiv{having 15 running} threads\arxiv{, i.e., using} 2 \arxiv{NUMA} nodes of the system. \arxiv{Second, we observe that \smartpq{}} successfully \arxiv{adapts} to the best-performing \algomode{} mode\arxiv{, and performs best in \emph{all} contention scenarios.} In Figure~\ref{fig:dynamic_total}, even when multiple features in the contention workload vary during the execution, \smartpq{} \arxiv{outperforms \emph{alistarh\_herlihy} and \nuddle{} by} \arxiv{1.87$\times$ and 1.38$\times$ on average, respectively. Note that any of the contention workloads evaluated in Figures~\ref{fig:dynamic_bench} and ~\ref{fig:dynamic_total} belongs in the training data set used for training our classifier.} \arxiv{Third, we note that the decision making mechanism of \smartpq{}} has very low performance overheads. \arxiv{Across all evaluated benchmarks, \smartpq{} achieves \emph{only up} to \arxiv{5.3\%} performance slowdown (i.e., when using a range of 50M keys in Figure~\ref{fig:dynamic_range}) over the corresponding baseline implementation (\emph{alistarh\_herlihy} priority queue). Note that since the proposed decision tree classifier has very low traversal cost (Section~\ref{sec:classifier}), we intentionally set a frequent time interval (i.e., one second) for calling the classifier, such that \smartpq{} detects the contention workload change \emph{on time}, and quickly adapts itself to the best-performing \algomode{} mode.  We also tried large time intervals, and observed that \smartpq{} slightly delays to detect the contention workload change, thus achieving lower throughput in the transition points. 

Overall, we conclude that \smartpq{} performs best across \emph{all} contention workloads and at \emph{any} point in time, and incurs negligible performance overheads over the corresponding baseline implementation.
}

\begin{table}
\begin{subtable}{1.01\linewidth}\centering
\resizebox{\textwidth}{!}{
{\begin{tabular}{r r r c c}
\toprule
\textbf{Time (sec)} & \textbf{Current Size} & \textbf{Key Range} & \textbf{Number of Threads} & \textbf{Insert - DeleteMin (\%)} \\
\midrule
0 & \textbf{1149} & \textbf{100K} & 50 & 75-25 \\
25 & \textbf{812} & \textbf{2K} & 50 & 75-25 \\
50 & \textbf{485} & \textbf{1M} & 50 & 75-25 \\
75 & \textbf{2860} & \textbf{10K} & 50 & 75-25 \\
100 & \textbf{2256} & \textbf{50M} & 50 & 75-25 \\
\bottomrule
\end{tabular}}
}
\caption{Varying the key range in the workload.}\label{tab:1a}
\end{subtable}%

\begin{subtable}{1.01\linewidth}\centering
\resizebox{\textwidth}{!}{
{\begin{tabular}{r r c c c}
\toprule
\textbf{Time (sec)} & \textbf{Current Size} & \textbf{Key Range} & \textbf{Number of Threads} & \textbf{Insert - DeleteMin (\%)} \\
\midrule
\midrule
0 & \textbf{1166} & 20M & \textbf{57} & 65-35 \\
25 & \textbf{15567} & 20M & \textbf{29} & 65-35 \\
50 & \textbf{15417} & 20M & \textbf{15} & 65-35 \\
75 & \textbf{15297} & 20M & \textbf{43} & 65-35 \\
100 & \textbf{15346} & 20M & \textbf{15} & 65-35 \\
\bottomrule
\end{tabular}}
}
\caption{Varying the number of threads that perform operations in the data structure.}\label{tab:1b}
\end{subtable}

\begin{subtable}{1.01\linewidth}\centering
\resizebox{\textwidth}{!}{
{\begin{tabular}{r r c c c}
\toprule
\textbf{Time (sec)} & \textbf{Current Size} & \textbf{Key Range} & \textbf{Number of Threads} & \textbf{Insert - DeleteMin (\%)} \\
\midrule
\midrule
0 & \textbf{1M} & 5M & 22 & \textbf{50-50} \\
25 & \textbf{140} & 5M & 22 & \textbf{100-0} \\
50 & \textbf{7403} & 5M & 22 & \textbf{30-70} \\
75 & \textbf{962} & 5M & 22 & \textbf{100-0} \\
100 & \textbf{8236} & 5M & 22 & \textbf{0-100} \\
\bottomrule
\end{tabular}}
}
\caption{Varying the percentage of \insrt/\delete{} operations.}\label{tab:1b}
\end{subtable}
\caption{\arxiv{Features of the contention workload for benchmarks evaluated in Figure~\ref{fig:dynamic_bench}}. We use bold font on the features that change in each execution phase.}\label{tab:features_dynamic_bench}
\end{table}

\begin{table}
\begin{subtable}{1.05\linewidth}\hspace{-8pt}
\resizebox{\textwidth}{!}{
{\begin{tabular}{r r c c c}
\toprule
\textbf{Time (sec)} & \textbf{Current Size} & \textbf{Key Range} & \textbf{Number of Threads} & \textbf{Insert - DeleteMin (\%)} \\
\midrule
\midrule
0 & \textbf{1M} & 10M & 57 & 50-50 \\
25 & \textbf{26} & 10M & \textbf{36} & \textbf{70-30} \\
50 & \textbf{12} & \textbf{20M} & 36 & \textbf{50-50} \\
75 & \textbf{79} & 20M & 36 & \textbf{80-20} \\
100 & \textbf{29K} & 20M & \textbf{50} & 80-20 \\
125 & \textbf{319K} & \textbf{100M} & 50 & \textbf{50-50} \\
150 & \textbf{13} & 100M & \textbf{57} & 50-50 \\
175 & \textbf{524K} & 100M & \textbf{22} & \textbf{100-0} \\
200 & \textbf{524K} & 100M & 22 & \textbf{50-50} \\
225 & \textbf{1142} & 100M & 22 & 50-50 \\
250 & \textbf{463} & \textbf{200M} & \textbf{57} & \textbf{0-100} \\
275 & \textbf{253} & 200M & 57 & \textbf{100-0} \\
300 & \textbf{33K} & \textbf{20M} & 57 & \textbf{0-100} \\
325 & \textbf{142} & 20M & \textbf{29} & \textbf{80-20} \\
350 & \textbf{25K} & 20M & 29 & \textbf{50-50} \\
\bottomrule
\end{tabular}}
}
\end{subtable}
\caption{\arxiv{Features of the contention workload for benchmarks evaluated in Figure~\ref{fig:dynamic_total}. We use bold font on the features that change in each execution phase.}}\label{tab:features_dynamic_total}
\end{table}

\section{Discussion \& Future Work}\label{sec:discussion}

In \arxiv{Section~\ref{sec:classifier}}, we assume that the \arxiv{contention} workload is known a priori \arxiv{to} extract the features needed for \arxiv{classification. To extract these features \emph{on-the-fly}, and \emph{dynamically} detect when contention changes, the main} structure of \smartpq, i.e., {\fontfamily{lmtt}\selectfont \textit{struct smartpq}}, \arxiv{needs to} be enriched with additional fields \arxiv{to} keep track of workload statistics \arxiv{(e.g., the number of completed \insrt/\delete{} operations, the number of active threads that perform operations on the data structure, the minimum and/or maximum key that has been requested so far). Active threads that perform operations on the data structure could \arxiv{atomically} update these statistics.} In frequent time lapses, either \arxiv{a} background thread or an active thread \arxiv{could} extract the features needed for classification based on the workload statistics, and call the classifier to predict if a transition \arxiv{to another \algomode{} mode} is needed. \arxiv{Finally, an additional parameter could be also added in \smartpq{} to configure} how often to collect workload statistics.

In our experimental evaluation, \arxiv{we pin server threads on a single NUMA node and client threads on all nodes. We have chosen to do so (i) for simplicity, given that this approach fits well with our microbenchmark-based evaluation, and (ii) because this is par with prior works on concurrent data structures~\cite{ascylib,ffwd,spraylist,blackbox,sagonas,Sagonas2016TheCA,numask,pim_cds,Choe2019Concurrent,Winblad2018Lock,adaptivepq,Siakavaras2017Combining,Guerraoui2016Optimistic,Howley2012ANon,Bronson2010APractical,Ellen-bst}. In a real-life scenario, where \smartpq{} is used as a part of an application,} client threads do \emph{not} need to be pinned in hardware \arxiv{contexts,} and they can be allowed to \arxiv{run in any core of the system}. However, for our approach to be meaningful server threads need to be limited \arxiv{on} a single NUMA node. This can easily be done by creating the server threads when \smartpq{} is initialized, and \arxiv{pinning} them to hardware \arxiv{contexts that are located at the same NUMA node}. In this case, \arxiv{server threads} are background threads that only accept and serve requests from various client threads, \arxiv{which} are part of the high-level application.

Finally, \arxiv{even though we focus on a microbenchmark-based evaluation to cover a \emph{wide variety} of contention scenarios, it is one of our future directions to explore the efficiency of \smartpq{} in real-life applications, such as} web servers~\cite{swift, web_servers}, graph traversal applications~\cite{sagonas,CLRS} and scheduling in operating systems~\cite{scheduler}. \arxiv{As future work, we also aim to investigate the applicability of our approach in other data structures, that may have similar behavior with priority queues (e.g., skip lists, search trees), and extend our proposed classifier (e.g., adding more features) to cover a variety of NUMA CPU-centric systems with different architectural characteristics.}

\setstretch{0.976}
\section{Related Work}
To our knowledge, this is the first work \arxiv{to propose an adaptive priority queue for NUMA systems, which performs best under \emph{all} various contention workloads, and even when contention varies over time. We briefly discuss prior work.}

\noindent\textbf{Concurrent Priority Queues.}
\arxiv{A large corpus of work proposes concurrent algorithms for} priority queues~\cite{lotan_shavit, linden_jonsson, sagonas, sundell, spraylist, adaptivepq, Wimmer_Martin, rihani, Brodal, Zhang,Sanders1998Randomized,Sagonas2016TheCA,Rab2020NUMA}, or generally \arxiv{for} skip lists~\cite{hotspot, fraser, fomitchev, herlihy, herlihy_art,dick, pim_cds,Pugh1990SkipLists,Choe2019Concurrent}. Recent works~\cite{lotan_shavit, linden_jonsson} designed lock-free priority queues that separate the logical and the physical deletion of an \arxiv{element to increase parallelism.} Alistarh et al.~\cite{spraylist} \arxiv{design} a relaxed priority queue, called \textit{SprayList}, in which \delete{} \arxiv{operation} returns with \arxiv{a} high probability, an \arxiv{element} among the \emph{first} $\mathcal{O}(p\log{}3p)$ \arxiv{elements of the priority queue}, where $p$ is the number of threads. \arxiv{Sagonas et al~\cite{Sagonas2016TheCA} design a contention avoiding technique, in which \delete{} operation returns the highest-priority element of the priority queue under \emph{low} contention, while it enables relaxed semantics when high contention is detected. Specifically, under high-contention a few \delete{} operations are queued, and later several elements are deleted from the head of the queue \emph{at once} via a combined deletion operation.} \arxiv{Heidarshenas et al.~\cite{Heidarshenas2020Snug} design a novel architecture for relaxed priority queues. These prior approaches are \notnuma{} implementations. Thus, in NUMA systems, they incur significant performance degradation in high-contention scenarios (e.g., \delete{}-dominated workloads in Section~\ref{sec:simbple_bench}). In contrast,} Calciu et al.~\cite{adaptivepq} \arxiv{propose a \emph{NUMA-friendly} priority queue employing the combining and elimination techniques.} Elimination allows \arxiv{the} complementary operations\arxiv{, i.e., }\insrt{} with \delete{}, to complete \emph{without} \arxiv{updating} the data structure, while combining is a \arxiv{technique similar to the delegation technique~\cite{Calciu2013Message,Klaftenegger2014Delegation,Lozi2012Remote,Petrovic2015Performance,Suleman2009Accelerating} of \nuddle{} and \ffwd{}~\cite{ffwd}.} 
Finally, Daly et al.~\cite{numask} propose an efficient technique to \arxiv{obtain} \numa{} skip lists\arxiv{, which however,} can only be applied to skip list-based data structures. \arxiv{In contrast,} \nuddle{} is a \arxiv{\emph{generic} technique to obtain} \numa{} data structures.

\noindent\textbf{Black-Box Approaches.}
Researchers have  proposed black-box \arxiv{approaches: any data structure can be made wait-free or \numa{} without effort or knowledge on parallel programming or NUMA architectures.} Herlihy \cite{wait_free} \arxiv{provides} a universal method to design wait-free implementations of any sequential object. However, this method remains impractical due to high overheads. Hendler et al.~\cite{flatcombining1} propose flat combining; a technique \arxiv{to} reduce synchronization \arxiv{overheads} by executing multiple client \arxiv{threads'} requests \emph{at once}. Despite \arxiv{significant} improvements~\cite{flatcombining2}, this technique provides high performance \emph{only} \arxiv{for a few} data structures (\arxiv{e.g.,} synchronous queues). \ffwd{}~\cite{ffwd} is \arxiv{black-box} approach, \arxiv{which} uses \arxiv{the delegation technique~\cite{Calciu2013Message,Klaftenegger2014Delegation,Lozi2012Remote,Petrovic2015Performance,Suleman2009Accelerating} to eliminate cache line} invalidation traffic \arxiv{over the interconnect}. However, \arxiv{\ffwd{} is} limited to single threaded performance. Calciu et al.~\cite{blackbox} propose a black-box technique, named \textit{Node Replication}, to obtain concurrent \arxiv{\numa{}} data structures. In \textit{Node Replication}, every NUMA node has replicas of the shared data structure, which are synchronized via a shared log. Although \ffwd{} and \textit{Node Replication} are \arxiv{generic} techniques \arxiv{to obtain \numa{} data structures, similarly to \nuddle, both of them use a \emph{serial asynchronized} implementation as the underlying \emph{base algorithm}. Thus, if they are used as the \numa{} \algomode{} mode in an adaptive data structure, which dynamically switches between a \notnuma{} and a \numa{} mode, both \ffwd{} and \emph{Node Replication} need a synchronization point between transitions to ensure correctness. Consequently, they would incur high performance overheads, when transitions between \algomode{} modes happen at a non-negligible frequency.}

\textbf{Machine learning in Data Structures.}
Even though machine learning is widely used \arxiv{to improve performance in many emerging} applications~\cite{athena, ppopp2018, cluster2018, benatia,Gronquist2021Deep,Memeti2019Using,KusumNG2016Safe,Michie1968Memo,Meng2019APattern,Dhulipala2020APattern,Sedaghati2015Automatic,Benatia2016SparseMF,Pengfei2020LISA}, there is a handful of works~\cite{smart,google} that \arxiv{leverage} machine learning to design \arxiv{\emph{highly-efficient} concurrent} data structures. Recently, Eastep et al.~\cite{smart} use reinforcement learning to \arxiv{on-the-fly tune} a parameter in \arxiv{the} flat combining technique~\cite{flatcombining1,flatcombining2}, which is used in skip lists and priority queues. \arxiv{Kraska et al.~\cite{google} demonstrate} that machine learning models can be trained to predict the position or existence of \arxiv{elements in key-value lookup sets}, and \arxiv{discuss} under which conditions \arxiv{learned index models can} outperform the traditional indexed data structures (e.g., B-trees). 
\section{Conclusion}

We propose \arxiv{\smartpq,} an adaptive concurrent priority queue for NUMA \arxiv{architectures}, which \arxiv{performs best under \emph{all} various contention scenarios, and even when contention varies over time. \smartpq{} has two key components. First, it is built on top of \nuddle{};} a generic \arxiv{low-overhead technique to obtain} efficient \numa{} data structures \arxiv{using \emph{any} concurrent \notnuma{} implementation as its backbone. Second, \smartpq{} integrates a lightweight decision making mechanism, which is based on a simple decision tree classifier, to decide when to switch between \nuddle{}, i.e., a \numa{} \algomode{} mode, and its underlying \emph{base algorithm}, i.e., a \notnuma{} \algomode{} mode.} Our evaluation \arxiv{over a wide range of contention scenarios demonstrates that} \smartpq{} switches between \arxiv{the two \algomode{} modes} with \arxiv{negligible overheads, and significantly outperforms prior schemes, even when contention varies over time. We conclude that \smartpq{} is an efficient concurrent priority queue for NUMA systems, and hope that this work encourages further study on adaptive concurrent data structures for NUMA architectures.}

\setstretch{0.97}
\SetTracking
 [ no ligatures = {f},
 outer kerning = {*,*} ]
 { encoding = * }
 { -40 } 

{

  \let\OLDthebibliography\thebibliography
  \renewcommand\thebibliography[1]{
    \OLDthebibliography{#1}
    \setlength{\parskip}{2pt}
    \setlength{\itemsep}{2pt}
  }
  \bibliographystyle{IEEEtranS}
  \bibliography{ref}
}

\end{document}